\documentclass[12pt]{article}
\usepackage{amsmath,amssymb,amscd}
\usepackage{graphicx,epsfig}

\begin{document}

\begin{center}

       {\bf \Large   The quark-gluon medium}               \\

\vspace{2mm}

{\bf I.M. Dremin\footnote{Email: dremin@lpi.ru},
A.V. Leonidov\footnote{Email: leonidov@lpi.ru}} \\

\vspace{2mm}

{\it Lebedev Physical Institute, Moscow 119991, Russia}\\

\vspace{5mm}

{\bf \Large Contents}

\end{center}
$\;\;\;\;$ {\bf \Large 1 Introduction.}

\medskip
{\bf \Large 2 The main experimental findings.}

\medskip
{\bf \Large 3 The microscopic description of the quark-gluon medium.}

\medskip
  3.1 Color Glass Condensate, Glasma, Quark-Gluon Plasma.
  3.2 Jet quenching and parton energy loss in dense non-abelian medium.

{\bf \Large 4 The macroscopic approach to the quark-gluon medium.}

  4.1 Equations of the in-medium QCD.
  4.2 The chromopermittivity.
  4.3 Classical polarization effects in the quark-gluon medium and its chromodynamical properties.
  4.4 Instabilities at high energies.
  4.5 Nonlinear effects and the color rainbow.
  4.6 Hydrodynamics (thermodynamical and mechanical properties of Quark-Gluon Plasma.
                                     \medskip

{\bf \Large 5 Some new possibilities at the LHC}

{\bf \Large 6 Conclusions.}

{\bf References}

\vspace{2mm}


\bigskip
\begin{abstract}
The properties of the quark-gluon medium observed in high energy nucleus-nucleus collisions are discussed. The main experimental facts about these
collisions are briefly described and compared with data about proton-proton collisions. Both microscopic and macroscopic approaches to their
description are reviewed. The chromodynamics of the quark-gluon medium at high energies is mainly considered. The energy loss of partons moving in
this medium is treated. The principal conclusion is that the medium possesses some collective properties which are crucial for understanding the
experimental observations.
\end{abstract}

{\bf A foreword}

\medskip

{\it It is a great honor to us to publish this paper in the volume dedicated to the memory of V.L. Ginzburg. His contribution to the condensed
matter physics crowned by the Nobel Prize is extremely important. Nowadays these ideas become actual for high energy physics. The properties of
the quark-gluon medium created in high energy nucleus collisions (often called the quark-gluon plasma) are intensively studied. The first steps of
chromodynamics of quark-gluon matter strongly remind those of electrodynamics of continuous media with some visible differences, however.}

\bigskip

\section{Introduction}

The topic of {\it the quark-gluon medium} is so widely spread nowadays that it deserves a series of monographs, even in spite of still being in
the actively developing stage, and can not be fully covered within a single review paper. That is why we mostly concentrate here on general ideas
about the evolution of the quark-gluon medium in high energy heavy ion collisions and on those properties of the medium which are revealed by the
energy losses of partons moving in it. They are described by the chromodynamics of the quark-gluon medium which is of the main topic of this
review.

Nevertheless, it would be too "ostrich-like" behavior not to mention the nearby problems and approaches. Therefore we include in the present
review brief discussions of some of them. In particular, the analogies and differences between electrodynamical and chromodynamical processes are
described. The extremely popular hydrodynamical approach has been widely discussed in many review papers. That is why we decided to give just a
short description of it here referring to the corresponding literature. The Anti de-Sitter/Quantum Chromodynamics (AdS/QCD) correspondence is
touched upon only slightly. Lattice results are mentioned very briefly too. They are mostly applicable to static properties of hadrons and are of
interest to us here in what concerns phase transitions. The very interesting ideas about possible local CP violation in heavy ion collisions are
briefly mentioned but not discussed in detail. The relation of the little bang in heavy ion collisions to the Big Bang in cosmology is another
topic of interest. However, it is out of the scope of our survey. A short overview of experimental results is presented in section 2. However, in
subsequent sections of this review only those experimental papers are cited  which are discussed in more details in connection with theoretical
considerations. We apologize to those whose work could not be completely reviewed due to the limited size of this paper.

The internal structure and interactions of particles and nuclei are the main goal of studies at accelerators. According to present knowledge
strong interactions are described by quantum chromodynamics (QCD). It describes quarks and gluons as elementary objects (partons) responsible for
the interaction.

At low energies the partons are confined inside strongly interacting particles (hadrons) and define their static properties. At large densities we
have the non-perturbative region of QCD describing the strongly interacting matter in thermal equilibrium at finite temperature.

The matter produced in collisions is surely different from the above-described one. Highly coherent parton configurations and strong internal
fields become especially important. By colliding two heavy nuclei at ultrarelativistic energies one expects to get a hot and dense internally
colored medium. It should exhibit some collective properties different from those seen at static conditions. Low-momenta modes in nuclear wave
functions can be described in terms of classical fields coupled to some high-momenta static color sources.

According to present knowledge, the collision process (energy evolution of all modes) proceeds through several stages. We discuss some of them at
the beginning of the review. The corresponding equations describing these processes are also discussed. At one of the stages quarks and gluons may
become deconfined inside some finite volume during a short time interval. According to present knowledge they form some ideal liquid. Its
collective properties are revealed by its mechanical motion as a whole described by hydrodynamics and by chromodynamical response to partons
penetrating it described by in-medium QCD.

At large transferred momenta the QCD coupling becomes smaller (the asymptotical freedom) and the perturbative approach starts being applicable.
Experiments on high energy particle collisions are needed precisely to make it possible to study the processes with high transferred momenta. In
accordance with the uncertainty principle it implies a possibility to learn the particle structure at ever smaller distances. In this way the
quark-gluon content of a particle, its scale dependence and the properties of the interaction region are studied.

Moreover, the hard process inside medium may induce its collective coherent response. That would allow to study medium properties like explosions
in earth depths serve to learn its structure. This is especially true at highest energies because the parton densities inside colliding strongly
interacting nuclei increase with energy. Therefore, the collective effects become pronounced and, in central heavy ion collisions, the long-range
correlations appear alongside with the short-range correlations .

To learn experimentally about the properties of the matter inside the interaction region one should study the energy losses of various trial
partons moving in it. This is analogous to studies of energy losses of electrons passing through the ordinary amorphous matter. Recent
experimental results of SPS (Super Proton Synchrotron) and RHIC (Relativistic Heavy Ion Collider) clearly show that the collective medium behavior
becomes crucial in heavy ion collisions. The radiation properties of partons are modified and energy spectra of final hadrons as well as their
correlations are changed relative to those in pp collisions. The induced coherent emission reveals the collective response of the medium to
partons penetrating it. The medium itself displays the collective motion. The hadrochemical content of produced particles changes. Both microscopic
and macroscopic QCD approaches to theoretical description of these findings are effective. Mechanical and thermodynamical properties of the medium
are studied in the hydrodynamical approach. We describe all of them in what follows together with discussion of corresponding experimental data.
Everywhere in this review we use the system of units in which $\hbar=c=1$.

\section{The main experimental findings}

The nucleus-nucleus (AA) collisions are studied experimentally at SPS and RHIC. The very first question addressed is the difference in AA and pp
processes at high energies. Various characteristics were measured and compared. They clearly show that nuclear collisions can not be considered as
an incoherent superposition of the nucleon-nucleon collisions. Some collective properties of the medium must be taken into account to explain the
difference.

{\bf The yield} of particles produced at rapidity $y$ and large transverse momenta $p_T$ is suppressed in AA relative to pp. This is demonstrated
by plotting the nuclear modification factor $R_{AA}$ for single-inclusive hadron cross section:
\begin{equation}
R_{AA}(y, p_T;b)=\frac {dN_{AA}/dydp_T}{N_{part}dN_{pp}/dydp_T}. \label{RAA}
\end{equation}
The factor $R_{AA}$ measures the deviation of AA reaction with the number of participant nucleons $N_{part}$ (determined by the nuclear overlap
function at impact parameter $b$) from an incoherent superposition of pp reactions for which $R_{AA}=1$. A similar expression can be written for
the dihadron correlations (usually called $I_{AA}$).

Both in particle and heavy ion collisions the production of the soft momentum particles with the transverse momenta $p_T<2$ GeV strongly
dominates. The corresponding spectrum decreases exponentially with transverse kinetic energy $\sqrt {p_T^2+m_i^2}-m_i$. The pion spectra are
flatter in the AA collisions than in the pp ones. This is even more true for kaons and antiprotons with the latter having the smallest slope.
These spectra are used to determine the kinetic freeze-out temperature and the dependence of the transverse flow velocity on the energy and
centrality (i.e. degree of collectivity of the interaction).

Accordingly, soft pions dominate multiplicity distributions. Their shape is approximately described by the negative binomial distributions both in
particle and heavy ion collisions at high energies albeit some deflections from them (predicted by QCD) are revealed by such correlation measures
as their moments.

The shape of the charged hadron transverse momentum spectra changes at $p_T \approx 3$ GeV from the exponential to the power-like as predicted by
the perturbative QCD. This power law tail affects very small yield of particles. The RHIC results on single particle inclusive distributions in
central Au-Au collisions at the energy of 200 GeV show the strong $p_T$-independent suppression, $R_{AA}\approx 0.2$, of the hadronic yield at
large $p_T>4$ GeV. The measurements covered the interval in $p_T$ up to 20 GeV. This striking deficit of the high transverse momentum particles
points out to the energy loss of partons in the medium. It corresponds to the so-called jet quenching effect with softened spectra of hadrons
produced by partons in the medium compared to the vacuum. Therefore, this suppression factor is a powerful tool to map out the density of the
medium.

The high $p_T$ suppression in Au-Au and Cu-Cu collisions is comparable at the same number of "participants" $N_{part}$ (nucleons which underwent
at least one inelastic interaction). The ratio $R_{AA}$ is closer to 1 for more peripheral collisions. It depends on the angular orientation of
the high-$p_T$ hadron with respect to the reaction plane (on the azimuthal angle). This dependence may be explained as due to the dependence of
the energy loss on the length traversed by the particle in the medium.

The ratio $R_{AA}$ does not depend on hadron identity at high $p_T$. In contrast to this, in electrodynamics the radiation of electrons is much
stronger than that of the muons. Interesting effects take place for suppression of the spectra of charmed mesons and charmonium. The former are
suppressed at the same level as pions while the latter is suppressed at the same level as at the SPS energy $\sqrt{s}=17\;{\rm GeV}$.

Another intriguing feature of the data is the fact that direct photons seem to be not suppressed in the region of $p_T$ from 4 GeV to 15 GeV while
at very large $p_T \sim 20\;{\rm GeV}$ preliminary data shows a suppression with $R_{AA}\approx 0.6$.

At smaller transverse momenta, $0.25<p_T<4$ GeV, the suppression effect is weaker, $R_{AA}$ even increases slightly at low $p_T<3$ GeV.  This is
qualitatively explained by the parton rescattering in the initial state which leads to the widening of the transverse momenta spectra (Cronin
effect). At the SPS energies the results refer only to this domain of transverse momenta with larger values of $R_{AA}$ than at RHIC. The SPS
value (Pb-Pb at $\sqrt s$=17.3 GeV) of $R_{AA}$ at $p_T=4$ GeV is much larger than at higher energies of RHIC exceeding 1 for $p_T>2$ GeV (strong
Cronin effect). That is awaited from the more important role of gluons as compared to quarks at RHIC ($C_A/C_V=9/4$). For collisions with smaller
overlap d-Au at 200 GeV the Cronin effect persists at least up to $p_T$=8 GeV. Experimental data show that the suppression of inclusive spectra
sets in somewhere in between  $\sqrt{s} \simeq 20 \; {\rm GeV}$ and $\sqrt{s} \simeq 60 \; {\rm GeV}$. The dominating process of soft particle
production is characterized by the scaling with the number of participants rather than by that in the number of binary collisions typical for hard
processes.

{\bf Correlations} among the secondary particles are the more sensitive instrument in distinguishing between models predictions than the single
particle inclusive data. Their studies revealed many exciting features. Most spectacular of them refer to jets produced in the processes in which
high $p_T$ particle is present. If a high-$p_T$ particle is chosen as a trigger and other hadrons are measured within a surrounding it
circle\footnote{Beside such cone algorithm, the recombination algorithms with iterative pairing of nearby particles and simple Gaussian filtering
are very widely used. These algorithms are helpful in the separation of jets from the background.} defined by the (pseudo)rapidity+azimuthal angles in 
the plane orthogonal to the momentum of the initial particle, then the Gaussian-like structure around the trigger (the near-side jet) is
usually seen in pp-reactions. Moreover, both two- and three-particle correlations show two back-to-back jet-like peaks (di-jet production).
Theoretically, jets are considered, at a first approximation, as survived remnants of hard-scattered quarks and gluons.

In AA collisions, jets are expected to widen in angles and momentum spectra of their particles are expected to soften. This phenomenon is termed
jet quenching. Quantitative analysis of jet quenching in heavy ion collisions requires model building. So far there are many uncertainties in
their predictions. Also, the biases are introduced by some experimental jet selection criteria. To be less dependent on these effects, different
triggers (e.g., $\gamma $ and $\pi ^0$) were used and the away-side hadron spectra were measured.

Correlations of non-photonic electrons from charm and bottom decays with hadrons are different for jets with $D$ and $B$ mesons.

Multiparticle correlations and fully (calorimetrically) reconstructed jets are the main goals of recent efforts in the studies of pp and AA
collisions. First results for fully reconstructed jets in pp, Cu-Cu, Au-Au collisions clearly demonstrate that jets are broadened in the
quark-gluon medium. In general, The color connections of jets to the bulk or, e.g., to beam remnants in W+jet events and the event-by-event
analysis of color flows are also important. The first experimental hints of intra-jet broadening due to such color flows in them are obtained.

The influence of jet quenching and background fluctuations on conclusions about jet fragmentation functions is actively studied now. Surprisingly
enough, no modification of them in AA collisions compared to the pp ones was observed at high $p_T$ in preliminary studies. Probably, it shows
that a biased jet population was selected (surface effect?) and the criteria of jet selection should be modified.

Even more surprising were observations of completely new topologies called ridge and double-humped events in AA collisions. In central collisions
the near-side jet happens to stand on the pedestal (ridge) with long extension in measured pseudorapidities up to $\vert \eta \vert =4$ and
steeply falling in the azimuthal direction. The presence of the ridge is independent of the jet peak. The characteristics of this peak are the
same as those of bulk particle production. However, the particle spectrum in the ridge is somewhat harder than that in the bulk. Moreover, this
structure is also seen in two-particle correlations without a jet trigger. These events include all charged hadrons with $p_T>150$ MeV. The
event-by-event three-particle correlation studies in central collisions at higher $p_T>1\;{\rm GeV}$ showed that the particles from the ridge are
uncorrelated in pseudorapidity between themselves and with particles in the peak. The production of the ridge is uncorrelated with the jet-like
particle production. The ridge disappears in the low-multiplicity peripheral collisions and at high $p_T$ of the trigger particle. Both the long pseudorapidity
extension of the ridge and large wide clusters found in two-particle correlations points out to the important role of the collective effects.

The away-side peak at $\Delta \phi = \pi $ is present in pp-collisions but is replaced by the broad away-side structure in Au-Au collisions. The
two symmetrical maxima (humps) are clearly seen in most central collisions at $\Delta \phi \approx \pi \pm 1.1$. They are resolved after the flow
effects are subtracted and contain hadrons with small energies. The peak positions are approximately independent of the transverse momenta of the
trigger and associated particles. The peak heights are slowly increasing with $p_T^{trig}$ at fixed $p_T^{assoc}$. The two humps seem to merge in
a single wide hump at large $p_T$ of the trigger ($6<p_T^{trig}<10$ GeV). It indicates on the reappearance of the away-side jet just in-between
these humps as it should be in the finite size medium where the parton with extra $p_T$ goes outside it in the form of a jet. The heavy quark jets
possess the similar qualitative features, albeit measured with the lower statistics. These peculiar features are seen both in two- and
three-particle correlations. They evidence for a conical emission pattern. The existence of these features is surely related to the collective
properties of the medium. In mid-central collisions the maximum position happens to be shifted for some special choices of the trigger location
relative to in- and out-of- planes directions.  This special feature can be explained by the effect of the wake (see below).

The associated yield of away-side hadrons is suppressed at large $p_T$ in Au-Au collisions relative to the pp ones, albeit less than the single
particle spectrum ($I_{AA}\approx 0.35 - 0.5$ while $R_{AA}\approx 0.2$).

Two-particle correlations between charged hadrons generated by final state Coulomb interactions provide valuable information. Also Bose-Einstein
(BE) correlations between pairs of identical bosons caused by quantum symmetrization of wave functions are studied with the help of a'la Hanbury
Brown-Twiss (HBT) interferometry. They are observed as an enhancement of pairs of same-sign charged pions at small relative momenta and tell us
about the space-time structure and evolution of the source. In central collisions at midrapidity the three radii $R_{long}, \, R_{out}, \,
R_{side}$ are usually reconstructed from measurements. They are directed, correspondingly, along the beam direction, that of the emission of the
pair ${\bf p}_T$ and the one perpendicular to these two. Within experimental errors all the three radii turn out to be equal and rather steeply
decreasing with transverse momenta. This contradicts the hydrodynamical predictions and is, therefore, called the "RHIC HBT puzzle". Moreover,
recent data show that the HBT characteristics of pp collisions behave in the similar way to those of AA which adds new questions to this puzzle.
The size of the correlated particle emission region increases with the particle multiplicity in the event. The number of the radii is larger (6)
for non-central collisions. They grow with the number of participants and exhibit oscillations as functions of the azimuthal angle. The growth
rate decreases with transverse momenta (as well as the radii themselves). The amplitude of oscillations is larger for more peripheral collisions
and vanishes in central collisions.

The modified shape of bosonic resonances created in nuclear collisions has been observed. In all cases the traditional Breit-Wigner form gets an
excess in its low-energy wing as is revealed in mass-correlations of the decay products.

{\bf Collective flows} of final particles present another interesting effect. These flows are generated by the internal pressure in the
quark-gluon medium. Their shapes are determined by spatial characteristics of initial overlap volumes and by interaction dynamics. The azimuthal
angle distribution of emitted particles can be written as
\begin{equation}
E\frac {d^3N}{d^3p}=\frac {1}{2\pi }\frac {d^2N}{dyp_Tdp_T}(1+
\sum _{n=1}^{\infty }2v_n\cos[n(\phi -\Psi _r)]),
\label{vAA}
\end{equation}
where $\Psi _r$ denotes the reaction (event) plane determined, independently for each component, from the flow itself: $v_n=<\cos [n(\phi -\Psi
_r)]>$ with average over all particles in all events as indicated by $<\;>$. The first harmonic coefficient of the Fourier expansion of azimuthal
distributions, $v_1$, describes the isotropic radial flow, and the second coefficient, $v_2$, characterizes the elliptic flow. To the anisotropic
flow there corresponds the azimuthal asymmetry of particle distribution with respect to the reaction plane defined by the beam direction and the
vector of impact parameter. The higher harmonics are also studied. The "non-flow" contributions due to jets, ridge, resonance decays, HBT effects
are usually subtracted.

The radial flow manifests itself in flattening (in AA collisions compared to the pp ones) of the spectra at small transverse kinetic energy which
is stronger for heavier particles. It scales with energy and the system size. This effect is more pronounced in central collisions between equal
spherical nuclei. The change of sign of radial flow of protons (the so-called "wiggle") as a function of rapidity for the most peripheral events
at SPS energies has been noticed. No wiggle is observed at RHIC. The coefficient $v_1$ stays negative (antiflow) and steadily increasing for all
pseudorapidities. This shows the early start of collective effects.

The most intensively studied elliptic flow is related, first of all, to the non-centrality of collisions (the spatial eccentricity of the reaction
zone) and to the rescattering of partons. The elliptic shape of the overlap region gives rise to different pressure in different directions and,
therefore, to non-zero $v_2$. Thus, the coefficient $v_2$ is sensitive to the system evolution at very early times. The integrated elliptic flow
increases with energy. This increase is stronger in more central collisions than in mid-central ones while the absolute values are smaller. The
value of $v_2$ depends on pseudorapidity with maximum at $\eta =0$ and increases with energy as well. It drops about a factor of two around $\eta
=3$. The value of $v_2$ is smallest for more central collisions. The dependence on the transverse momentum has been measured up to $p_T=12$ GeV.
It shows the maximum about $p_T=3$ GeV with $v_2$ becoming as large as 0.2 and then some saturation or slow decrease. At $p_T<2$ GeV $v_2$ is
characterized by the scaling with the transverse kinetic energy $m_T-m$ and at $2<p_T<4$ GeV -- with the number  of constituent quarks. The
relation of $v_2$ to the spatial eccentricity of an event depends on the number of participants (system size). The coefficient $v_2$ has been
measured for different particle species. The baryonic elliptic flow saturates at somewhat higher $p_T$ and larger $v_2$ values than for mesons.
The value of $v_2$ is close to 0 for $J/\psi $ in Au-Au collisions at 200 GeV.

{\bf Enhancement} of the yield of various particle species at high $p_T$ in AA collisions as compared to the very strong suppression of pions has
been observed. Thus the hadrochemical compositions in AA and pp collisions are different. This is considered as a manifestation of the quark-gluon
plasma (or, more generally, of the prehadronic state) effects. Neutral pions and $\eta $-mesons at high $p_T>2$ GeV are strongly (about 5 times)
suppressed in central AA collisions relative to pp processes. The suppression of $\phi $ mesons is the same at high $p_t>5$ GeV but smaller in the
intermediate $2<p_T<5$ GeV range. No suppression is seen for photons in this kinematic range. The proton production is enhanced at intermediate
$p_T\approx 2-5$ GeV so that the ratio $p/\pi $ becomes close to 1. Quite surprising is the fact that in AA at RHIC antiprotons are as abundant as
negative pions for $p_T>2$ GeV. That is probably related to the collective transverse flow. In particle collisions the strange particles
production at high $p_T$ is strongly suppressed compared to pions (3 - 4 times) but it is enhanced by about the factor of two in AA processes so
that the suppression factor becomes about 0.6 - 0.7. The enhancement is larger for particles with higher strangeness content such as $\Xi $,
$\Omega$. The overall fraction of strange particles is about twice as high in heavy-ion collisions as compared to the elementary particle
collisions. The similar relatively small suppression factor was measured for charmed quarks ($J/\psi $). The upper limit for bottom ($\Upsilon $)
suppression of 0.64 was recently obtained. A strong suppression of the high-$p_T$ electrons originating from the decays of charm and beauty
hadrons indicates the same value of suppression for heavy and light quarks in the quark-gluon medium. This shows that the initial stage of
creation of highly virtual quarks is more important than their subsequent mass difference. The change in the hadrochemical content  of collision
products with high $p_T$ may reflect a difference in in-medium interactions of the species and asks for taking into account the collective
(non-local) properties of the medium (with possible diffusion before hadronization).

Remarkably enough, at very low $p_T<0.5$ GeV of secondary particles their yields seem to be the same in pp, AA and even $e^+e^-$ processes (albeit
different for different species) and independent of the primary energy as is expected from universality of coherent processes for long-wavelength
gluons.

The above statements can be illustrated by numerous figures which can be found in experimental presentations and more specialized review papers.
We will only show some of them to elucidate theoretical argumentation. That would allow to keep the reasonable size of this article.

All these observations are closely related to the general physical process of the energy loss of partons during nuclei collisions which is to the
large extent the topic to which the present review is devoted to.

\section{The microscopic description of the quark-gluon medium}

The microscopic QCD description of all processes with thousands of particles produced is surely impractical in the whole phase space region. Such
a description may be applicable to the relatively rare high-$p_T$ subprocesses where the QCD coupling constant becomes smaller. For soft partons
and hadrons the rescattering and hadronization effects are too strong to leave many traces of the primary inelastic process. The macroscopic
approach (including the statistical one) is more suitable for description of purely collective medium effects.

One can nevertheless follow general evolution of fields during heavy ion collisions and reveal typical correlation patterns by applying main
principles of QCD. In this context, the present paradigm, which we describe below, pictures the transition from the Color Glass Condensate (CGC)
to Glasma eventually evolving into Quark Gluon Plasma (QGP) and subsequent hadronization .

The energy loss of partons in the quark-gluon medium is the main source of experimental information about its properties during these stages. As
in electrodynamics, it may be separated in the two categories.

The loss due to the change of the velocity vector of a parton such as elastic scattering, bremsstrahlung and synchrotron radiation is usually
treated microscopically. All these phenomena result from the {\it short-range} response of the parton to the impact of the matter fields. Elastic
scattering does change the parton energy due to the recoil effect and deflects it and thus changes the energy flow in the initial direction. At
high energies this process is less probable than the emission of gluons (bremsstrahlung) due to the nearby collisions with the matter
constituents. The medium structure imposes some restrictions on the coherence properties of the radiation process and on the effective radiation
length (in analogy to the Ter-Mikhaelyan and Landau-Pomeranchuk effects in electrodynamics). In the general case, the intensity of the radiation
depends on the relation of the path length of the parton in the medium, its mean free path (depending on the distance between the scattering
centers and on the cross section) and the formation length of emitted radiation. The synchrotron radiation of gluons induced by the curvature of
parton trajectory in chromomagnetic fields may become important for strong enough fields. These processes are considered in the section 3.

A principally different source of energy losses is connected with the medium polarization by the propagating parton. Here one speaks about the
collective response of the medium related to the non-perturbative {\it long-range} interconnection of its constituents described by its
chromopermittivity. This is treated macroscopically within the in-medium QCD. The corresponding effects are analogous to the Cherenkov radiation,
the wake and the transition radiation. In these processes one can, in the first approximation, neglect the change of the velocity vector of a
parton. The macroscopical aspect of the problem is sometimes ignored in the review papers and we will discuss it in more details in section 4.

Let us note that the total energy loss is the sum of the losses due to short-range and long-range interactions.

\subsection{Color Glass Condensate, Glasma, Quark-Gluon Plasma}

Is it possible to work out a quantitative description of multiparticle production in hadronic and nuclear collisions within QCD? An ambitious
attempt of answering this question is provided by the paradigm of Color Glass Condensate (CGC), see e.g. the reviews \cite{CGC1,CGC2,CGC3,CGC4}.
In the words of one of the principal creators of this paradigm, "Color Glass Condensate is a universal form of matter which controls the high
energy limit of all strong interaction processes and is the part of the hadron wavefunction important at such energies" \cite{M03}.

The strength of the CGC approach to the physics of high energy scattering is in providing a unifying concept for such important but previously
separately existing notions as soft (wee) and constituent partons, stringy and jetty particle production, structure functions and limiting
fragmentation, etc.

Multiparticle production in QCD is generated by low $x$ partons in the nuclear wavefunctions (CGC). Chromoelectric and chromomagnetic fields in
the medium under consideration are the strongest fields in nature. When two "pancakes" of CGC collide, they form matter with very high density
called Glasma. It preceeds the Quark-Gluon Plasma (QGP). Glasma consists of color flux tubes localized in the transverse plane and stretching
between the valence color degrees of freedom.  These tubes generate long-range correlations responsible for flat distributions in pseudorapidity
$\Delta \eta$. The existence of long-range (pseudo)rapidity correlations is related to the early times of Glasma
formation in nuclear collisions $\tau \leq \tau_{\rm out} \exp (-\vert y_a-y_b \vert/2)$, where $\tau_{\rm out}$ is the proper time of production
of the particles $a$ and $b$ with the rapidities $y_a$ and $y_b$. The radial flow collimates particles in the direction of the flow and generates
azimuthal correlations but does not affect the distribution over $\Delta \eta$. In particular, the ridge could be explained as the combination of
the initial state long-range rapidity correlations and the final state radial flow. Therefore combined elements of the both micro-and macro-
approaches are necessary to explain the ridge phenomenon.

Color Glass Condensate is an effective theory describing hadronic and nuclear wavefunctions in the regime where the main contribution to the
cross-section is coming from their multiparton Fock components, i.e. in essence from the dense gluon matter. The key idea behind the physical
picture of CGC can be formulated as follows. Let us consider a fast hadron (nucleus) moving along the $z$ axis with the big longitudinal
light-cone momentum $P^+$. The modes in the projectile are classified by their relative yields of longitudinal momenta $x=p^+/P^+$. Then, from
uncertainty relation, the characteristic longitudinal extension of these modes is $\Delta x^- \sim 1/(xP^+)$ and their characteristic lifetime is
$\Delta x^+ \sim 1/p^-  \backsimeq 2xP^+/m^2_\perp$ \footnote{The light-cone coordinates $x^+$ and $x^-$ are defined below in equation
(\ref{lccoor})}, where $m^2_\perp$ is some characteristic scale of transverse momentum. We see that the hard constituent modes, for which $x \sim
1$, are localized at small distances and have large lifetimes, while the soft wee modes, for which $x \ll 1$, are spread over large longitudinal
distances and are short-lived. From the data on structure functions from deep inelastic scattering we know that the number of wee partons, most
importantly gluons, is rapidly growing with $1/x$, \; $xG(x,Q^2) \sim (1/x)^\kappa$ with $\kappa \simeq 0.3$, so that these modes are
characterized by large occupation numbers. For such modes it is natural to replace the language of particles by that of fields, and, in turn,  for
the coherent multiparticle states the appropriate description is that in terms of {\it classical} fields.

These considerations lie at the origin of the McLerran-Venugopalan model of nuclear light-cone wavefunctions \cite{MV1,MV2,MV3} in which, in the
framework of QCD, the constituent and wee partons are integrated in the unifying picture in which strongly localized slowly evolving constituent
hard modes are considered as the {\it source} for the delocalized soft wee gluon field for which one has to take into account nonlinear effects
leading, in particular, to the saturation of the gluon distribution, see below.

Let us introduce the variables to be used in this section. Most of the formulae are written in terms of light cone coordinates. For the
four-vector $x^\mu = (t,{\bf x}_\perp,z)$ the definition of the $\pm$-components reads
\begin{equation}\label{lccoor}
x^\pm  = \frac{t \pm z}{\sqrt{2}}
\end{equation}
The above-described momentum components are described analogously. We shall also make an extensive use of the $(\tau,\eta)$ coordinate system
\begin{equation}
 \tau = \sqrt{\;2 x^+ x^-}, \,\,\, \eta = \frac{1}{2} \ln \left( \frac{x^+}{x^-} \right),
\end{equation}
where $\tau$ is the proper time and $\eta$ is the spatial rapidity.

\subsubsection{McLerran-Venugopalan model. Saturation.}\label{mvm}

To illustrate the machinery of the McLerran-Venugopalan model, let us consider
a calculation of the momentum space density of gluon modes
\begin{equation}\label{corfun}
 \frac{dN}{d^3k} = \langle a^{i \dag}_c (x^+,{\vec k}) a^i_c (x^+,{\vec k}) \rangle
 = \frac{2k^+}{(2\pi)^3} \langle A^i_a(k,x^+) A^i_a(-k,x^+)  \rangle,
\end{equation}
from which one can calculate the observable gluon structure function
\begin{equation}\label{strfun}
xG(x,Q^2)=\int^{Q^2} d^2 {\bf k}_\perp  \left. k^+ \frac{dN}{dk^+ d^2 {\bf k}_\perp} \right \vert_{x=k^+/P^+} .
\end{equation}
As we are interested in the high density regime, the gauge potentials $A^i_a(k,x^+)$ in (\ref{corfun}) correspond to soft wee  gluon fields.

To calculate the momentum space gluon density in (\ref{corfun}) one needs to specify the gluon fields $A^i_a(k,x^+)$ and the nature of averaging.
The first task is achieved by assuming that $A^i_a(k,x^+)$ satisfy the {\it classical} Yang-Mills equations with the source originating from hard
constituent modes
\begin{equation}\label{cymeq}
\left[ D_\mu,F^{\mu \nu}\right] = \delta^{\mu +} \rho_1({\bf x}_\perp,x^-) ,
\end{equation}
where $F^{\mu\nu}=\left[D^\mu,D^\nu \right]$ is the non-abelian field strength and $D^\mu = \partial^\mu - ig A^\mu$. The solution of
(\ref{cymeq}) is conveniently written in the light-cone gauge $A^+=0$. It is easy to verify that for the static $x^+$-independent solution one has
$A^-=0$ so that the solution of (\ref{cymeq}) is purely transverse:
\begin{equation}\label{lw}
A^i ({\bf x}_\perp,x^-) =  \frac{i}{g} U({\bf x}_\perp,x^-) \partial^i U^\dag({\bf x}_\perp,x^-) ,
\end{equation}
where
\begin{equation}\label{poe}
U({\bf x}_\perp,x^-) = P \exp \left \{  ig \int_{-\infty}^{x^-} dy^- \alpha({\bf x}_\perp,x^-)  \right \} ,
\end{equation}
and $\alpha({\bf x}_\perp,x^-) =  -\rho({\bf x}_\perp,x^-)/\nabla^2_\perp$ is the solution of
(\ref{cymeq}) in the covariant gauge. The solution (\ref{lw}),(\ref{poe}) is a nonlinear functional of the
color charge density $\rho({\bf x}_\perp,x^-)$ and, being exact, is nevertheless formal.

The second crucial step is related to the above-mentioned specification of the nature of averaging in (\ref{corfun}). The idea of
\cite{MV1,MV2,MV3} was that for a large nucleus and the transverse resolution of the probe $\vert \, \Delta {\bf x}_\perp \vert \ll 1/\Lambda_{\rm
QCD}$, where $\Lambda_{\rm QCD}$ is the fundamental scale of strong interactions, $\Lambda_{\rm QCD} \sim 100 \; {\rm MeV}$, the color density
$\rho({\bf x}_\perp,x^-)$ is, on the event-by-event basis, essentially a random function. In the simplest setting where the hard modes are
constituent quarks the static color density is just the total color charge of all constituent quarks, that happen to lie in the tube under
consideration.  An averaging procedure in (\ref{corfun}) is thus specified by some functional $W_{\Lambda^+} \left[ \rho \right]$, where
$\Lambda^+$ is the scale separating hard modes from the soft ones. A simple but quite realistic averaging is provided by the Gaussian ensemble
such that \footnote{For a detailed analysis supporting this assumption see \cite{JV04}.}
\begin{equation}\label{rhocor}
\langle \rho^a({\bf x}_\perp,x^-) \rho^b({\bf y}_\perp,y^-)\rangle_{W_{\Lambda^+}}=
g^2 \mu^2_A \delta^{ab} \delta^2 ({\bf x}_\perp-{\bf y}_\perp)
\delta(x^--y^-)). \nonumber
\end{equation}
In the simplest case where the sources are constituent quarks the factor $\mu^2_A$ setting the scale for the transverse density of color charge is
simply $\mu^2_A=g^2A/(2\pi R_A^2)$ and is thus proportional to $A^{1/3}$ (see e.g. a detailed derivation in \cite{CGC1}). The color charge density
$\mu_A$ is directly related to the physical saturation scale $Q_s$ to be defined below, $Q_s \thickapprox 0.6 \; g^2 \mu_A$ \cite{L08}.

It turns out convenient to present the expression for the gluon density (\ref{corfun}) resulting after averaging over the stochastic sources
$\rho$ with the above-defined Gaussain weight for the gluon density in transverse space $\varphi(x,{\bf k}_\perp^2)$\footnote{The convenience of
introducing $\varphi(x,{\bf k}_\perp^2$) lies in particular in the fact that the quantum evolution of gluon distributions considered below is most
naturally written precisely in terms of this so-called unintegrated structure function.}
\begin{equation}
 \varphi(x,{\bf k}_\perp^2) = \frac{4\pi^3}{N^2_c-1} \frac{1}{\pi R_A^2} \frac{d^3 N}{d \ln(1/x) d^2 {\bf k}_\perp} .
\end{equation}
The characteristic properties of the resulting distribution can conveniently be presented by writing down its asymptotic expressions\footnote{For
details of this calculation see, e.g., \cite{CGC1}.} at ${\bf k}_\perp^2 \to 0$ and ${\bf k}_\perp^2 \to \infty$:
\begin{equation}\label{asymp}
\begin{cases}
\left. \varphi(x,{\bf k}_\perp^2) \right \vert_{{\bf k}_\perp^2 \to \infty} & \backsimeq
\frac{1}{\alpha_s N_c} \frac{Q^2_s}{{\bf k}_\perp^2} \equiv \frac{\mu^2_A}{{\bf k}_\perp^2},\\
\left. \varphi(x,{\bf k}_\perp^2) \right \vert_{{\bf k}_\perp^2 \to 0} & \backsimeq
\frac{1}{\alpha_s N_c} \ln \frac{Q^2_s}{{\bf k}_\perp^2}.
\end{cases}
\end{equation}
The most important ingredient of (\ref{asymp}) is the so-called saturation scale $Q^2_s \sim A^{1/3}$. The $A$-dependence of $Q^2_s$ is directly
related to that of $\mu^2_A$.

The remarkable phenomenon of flattening of the gluon distribution at small transverse momenta seen in (\ref{asymp}) is called saturation. We see
that saturation is characteristic of the regime in which the gluon density $\varphi \sim 1/\alpha_s$ is parametrically large. Another crucial
observation which is of great significance for developing the quantum theory of gluon distributions is that at large $A$ the saturation scale
becomes large, $Q^2_s \gg \Lambda^2_{\rm QCD}$. Since the saturation momentum $Q_s$ sets the scale for all the transverse momenta in the problem,
for asymptotically large nucleus the theory is in the weak coupling regime.

\subsubsection{Glasma}

In the paragraph 3.1.1 we have described, in the framework of color glass condensate theory, the distribution of gluons inside the nucleus. With
this knowledge at hand we can consider a problem of particle production in high energy nuclear scattering. The main idea of CGC-motivated approach
to this problem is that, if gluon distributions in nuclei before the collision are most naturally described in terms of fields, it makes full sense
to describe the energy and particle flux created at the early stage of these collisions also in terms of gluon field. In the tree-level
approximation this amounts to solving the Yang-Mills equations with two external currents $\rho_{1,2}({\bf x}_\perp,x^-)$ representing color
distributions of the two incident nuclei propagating along $x^\pm=0$:
\begin{equation}\label{ymts}
\left[ D_\mu,F^{\mu \nu}\right] = \delta^{\mu +} \rho_1({\bf x}_\perp,x^-) + \delta^{\mu +} \rho_2({\bf x}_\perp,x^-).
\end{equation}
If we are interested in calculation of the gluon production in the central region, where one can assume that their rapidity distribution is
approximately flat, it is natural to look for the boost-invariant solution of (\ref{ymts}) in the form  \cite{KMW95}:
\begin{eqnarray}
A^i & = & \theta(-x^+)\theta(x^-) A^i_{(1)}+\theta(x^+)\theta(-x^-) A^i_{(2)}+\theta(x^+)\theta(x^-) A^i_{(3)} \nonumber \\
A^\eta & = & \theta(x^+)\theta(x^-) A^\eta_{(3)}, \nonumber
\end{eqnarray}
Here $A^i_{(1,2)}$ are the gluon fields  of the two incident nuclei before the collision described by (\ref{lw}) and $ A^i_{(3)}$ and
$A^\eta_{(3)}$ are the gluon fields created after the collision that is assumed to take place at $\tau=0$. Taken at the proper time $\tau=0^+$,
these fields describe the initial conditions for subsequent evolution of dense matter created in high energy nuclear collisions.

It is straightforward to verify that to ensure the  matching of the solution of (\ref{ymts}) at $\tau=0$ the following relations should take
place:
\begin{eqnarray}\label{bc}
A^i_{(3)} \vert_{\tau=0} & = & A^i_{(1)} + A^i_{(2)}, \nonumber \\
A^\eta_{(3)} \vert_{\tau=0} & = & \frac{ig}{2} \left[A^i_{(1)},A^i_{(2)}\right].
\end{eqnarray}
A crucial observation \cite{FKL06,LM06} following from the boundary conditions (\ref{bc}) is that at $\tau=0^+$ there exist only longitudinal
fields
\begin{eqnarray}\label{lf}
E^z & = & ig \left[A^i_{(1)},A^i_{(2)}\right], \nonumber \\
B^z & = & ig \epsilon^{ij }\left[A^i_{(1)},A^j_{(2)}\right],
\end{eqnarray}
in which the energy created in the collision is stored. Thus there takes place a full transformation of the structure of the solution, i.e. the
transition from the two transverse fields $A^i_{(1,2)}$, described in the framework of CGC at $\tau=0^-$, to the purely longitudinal chromoelecric
and chromomagnetic fields immediately after the collision at $\tau=0^+$ called glasma. Let us stress that, as obvious from (\ref{lf}), the effect
is purely non-abelian. As the only scale in the problem is the saturation scale $Q_s$, the correlation length in the impact parameter plane is
$R_\perp \sim 1/Q_s$. Thus, the physics is that of the longitudinal fields characterized by the transverse correlation radius $R_{\perp}$, i.e.
with a set of the chromoelectric and chromomagnetic cylindrical tubes. It is at this point where the CGC description matches the old paradigm
of high energy hadronic collisions, the Lund string model. An important new ingredient, however, is the presence of the {\it chromomagnetic} flux
tubes. The very fact that ${\vec E} \cdot {\vec B} \neq 0$ opens up a principal possibility of the local CP violation \cite{LM06}.

In the course of subsequent evolution the initial configuration of longitudinal fields is undergoing a partial transformation into that also
containing transverse field and at $\tau \sim 1/Q_s$ the energy stored in longitudinal and transverse modes is approximately equal \cite{LM06}.

The expressions for the number of initially produced gluons and their transverse energy can be written from dimensional reasons\footnote{Because
of the ultraviolet divergence at $\tau=0$ these quantities can meaningfully be computed only at nonzero $\tau>0$ \cite{L06}.}:
\begin{equation}\label{net}
\left. \frac{dN}{d\eta} \right \vert_{\eta=0} = c_N \frac{\pi R_A^2 Q^2_s}{\alpha_s (Q_s)}, \;\;\;\;\;\;\;
\left. \frac{dE_\perp}{d\eta} \right \vert_{\eta=0} = c_E \frac{\pi R_A^2 Q^3_s}{\alpha_s (Q_s)},
\end{equation}
where the coefficients $c_N \approx 0.1/4 \pi$ and $c_E \approx 0.05/4 \pi$, (see e.g. \cite{V05} and references therein), are determined from
numerical solutions of (\ref{ymts}).

In particular, using the value of $Q_s$ determined from the analysis of deep inelastic scattering at HERA for $x_{\rm eff} \sim 0.01$ as
appropriate for nuclear scattering at RHIC, $Q_s({\rm RHIC})\backsimeq 1.2 \; {\rm GeV}$, one gets from (\ref{net}) the rapidity density of the
number of gluons and of the transverse energy:
\begin{equation}\label{nexp}
\left. \frac{dN}{d\eta} \right \vert_{\eta=0} \approx 1100\;, \;\;\;\;\;
\left. \frac{dE_\perp}{d\eta} \right \vert_{\eta=0} \approx 500 \; {\rm GeV},
\end{equation}
which agrees with the experimental data. This can be interpreted as an indication of almost perfectly ideal isentropic evolution from the initial
glasma phase, through the intermediate QGP stage, to that of the flux of final hadrons.

\subsubsection{Glasma instabilities}

In the section 3.1.2 we have considered a boost-invariant solution of the classical Yang-Mills equations (\ref{ymts}) describing the conversion of
two initial gluon fields inside the incident nuclei into the physical fields created in the collision. The latter can be described as a set of
chromoelecetric and chromomagnetic flux tubes having the radii of order of inverse saturation momentum $Q_s$ stretched between the receding
sources.

In general, one expects that classical solutions of Yang-Mills equations are unstable with respect to quantum fluctuations. For example, constant
chromoelectric fields are unstable with respect to Schwinger gluon pair production and static chromomagnetic fields are characterized by the
Nielsen-Olesen instability. A tree-level iterative calculation of gluon production in collisions of the non-abelian charges in \cite{MMR98} has
also demonstrated the instability of the approximate tree-level solution with respect to higher order corrections.

Thus it is probably not too surprising that the evolving glasma fields are unstable with respect to rapidity-dependent fluctuations. Such
fluctuations are inevitable when one takes into account quantum fluctuations. An explicit calculation of the wavefunction of two colliding nuclei
demonstrating this effect was done in \cite{FGM07}.

In describing the corresponding instabilities it is convenient to consider two regimes:
\begin{itemize}
\item{Small proper times $\tau < 1/Q_s$. This is a regime in which instabilities are those related to fields.}
\item{Intermediate times $\tau > 1/Q_s$. This regime can be described in dual terms of fields and particle
kinetics, see \cite{MS04}.}
\end{itemize}

The analysis of these instabilities was brought to a new level in the all-order numerical calculations made in \cite{RV061,RV062} in which it was
shown that the amplitudes of the soft modes grow like $\exp (\sqrt{\tau})$. An analytical analysis of instabilities with respect to quantum
fluctuations in the flux tube picture was performed in \cite{FI08}.

The physical origin of these instabilities is still under debate.

The closest known analogy is the non-abelian Weibel instability \cite{mr881, mr882, ALM03} with plasma instabilities of both electric and magnetic
type developing in non-equilibrium plasmas with inhomogeneous distributions of particles in momentum space. The analogy to glasma is very natural
because,  as described in the previous paragraph, the initially created configuration of glasma fields is maximally anisotropic (the initial
transverse momentum is equal to zero).

The analysis carried out in \cite{ALM03} have lead to the complete revision of the  previously existing picture of particle production in nuclear
collisions given in the framework of the so-called "bottom-up" scenario of thermalization of the glue initially freed at the momentum scale $Q_s$
described in \cite{BMSS01}. The reason for revision was that instabilities are much more effective in producing soft gluons and driving the system
towards isotropy faster than the bremsstrahlung mechanism considered in \cite{BMSS01}. The new versions of the "bottom-up" scenario include, in
particular, mechanisms of cascade turbulent-like multiplication of gluons \cite{AM06,MSW07,K08}.

\subsubsection{Quantum evolution. High-energy factorization.}

The McLerran-Venugopalan model is by construction adequate for moderate values of fractional light-cone momentum $x=k^+/P^+$. The quantum
corrections to the tree-level correlators like (\ref{corfun}) involve soft gluon emission and virtual gluon loops. Both effects bring in
contributions which paramerically are of the order of $O(\alpha_s \log(1/x)$. As physical quantities can not depend on the arbitrary separation
scale of the soft and hard modes $\log(\Lambda^+/P^+)$, there arises a Wilsonian renormalization group resumming such contributions. For dense
gluon systems this resummation of small-$x$ enhanced terms has to include contributions to all orders in the gluon density. The
logarithmic nature of quantum corrections makes it natural to use the rapidity variable $Y=\log(1/x)$. In these terms one can, for example,
interpret the gluon two-point function (\ref{corfun}) as an average of the product of classical fields taken at some specified rapidity scale:
\begin{equation}
<A A>_{Y} = \int {\cal D} \rho \,W_{Y}[\rho]\, A_{\rm cl.}(\rho)\,A_{\rm cl.}(\rho),
\end{equation}
The Wilsonian RG equation (the Jalilian-Marian-Iancu-McLerran-Leonidov-Kovner (JIMWLK) evolution equation \cite{JIMWLK1,JIMWLK2,JIMWLK3}) is most
naturally written as an equation for the functional $W_Y[\rho]$. As shown in \cite{W02} it can be cast in an elegant Hamiltonian form:
\begin{equation}\label{JIMWLK}
\frac{\partial W_Y [\rho]}{\partial Y}={\cal H}[\rho] \; W_{Y} [\rho],
\end{equation}
where ${\cal H}[\rho]$ is the JIMWLK Hamiltonian
\begin{equation}\label{HJIMWLK1}
{\cal H}[\rho] = \frac{1}{2} \int_{ {\bf x}_\perp \; , {\bf y}_\perp } \;
\chi ( {\bf x}_\perp, {\bf y}_\perp )  \; \frac{\delta^2}{ \delta \rho( {\bf x}_\perp ) \delta \rho( {\bf y}_\perp ) }
+\int_{ {\bf x}_\perp } \; \sigma( {\bf x}_\perp ) \; \frac{\delta}{\delta \rho( {\bf x}_\perp ) }.
\end{equation}
In equation (\ref{HJIMWLK1}) the functions $\chi ( {\bf x}_\perp, {\bf y}_\perp )$ and $\sigma ( {\bf x}_\perp )$ are explicitly calculable
\cite{JIMWLK1,JIMWLK2,JIMWLK3} and correspond to the effects due to emission of real and virtual gluons respectively.

A remarkable identity relating the real and virtual kernels  $\chi ( {\bf x}_\perp, {\bf y}_\perp )$ and $\sigma ( {\bf x}_\perp )$ discovered in
\cite{W02} allows to rewrite (\ref{HJIMWLK1}) in terms of  the real kernel $\chi ( {\bf x}_\perp, {\bf y}_\perp )$ only:
\begin{equation}\label{HJIMWLK2}
{\cal H}[\rho] = \frac{1}{2} \int_{ {\bf x}_\perp \; , {\bf y}_\perp } \
\frac{\delta}{\delta \rho( {\bf x}_\perp )} \; \chi ( {\bf x}_\perp, {\bf y}_\perp )  \; \frac{\delta}{\delta \rho( {\bf y}_\perp )}
\end{equation}
The explicit expression for $\chi ( {\bf x}_\perp, {\bf y}_\perp )$ involves the path-ordered exponentials $U({\bf x}_\perp)$ defined in equation
(\ref{poe}):
\begin{eqnarray}
 \chi ( {\bf x}_\perp \; ,  {\bf y}_\perp ) & = & \int \frac{d^2 {\bf z}_\perp}{4\pi^3}
 \frac{({\bf x}_\perp-{\bf z}_\perp)({\bf y}_\perp-{\bf z}_\perp)}{({\bf x}_\perp-{\bf z}_\perp)^2({\bf y}_\perp-{\bf z}_\perp)^2} \nonumber \\
 & & \left[ \left(1-U^\dag ({\bf x}_\perp) U ({\bf z}_\perp) \right) \left(1-U^\dag ({\bf z}_\perp) U ({\bf y}_\perp) \right)\right]
\end{eqnarray}

Until recently the usage of the JIMWLK evolution equations and their simplified versions, the BK equations \cite{BK1,BK2} corresponding to
retaining only the two-point functions in the infinite hierarchy generated by (\ref{JIMWLK}), was restricted to one-source problems like deep
inelastic scattering at high energies and quantities like gluon structure functions. In the major recent development \cite{HEF1,HEF2,HEF3} the
formalism allowing systematic computation of quantum corrections to the two-source glasma problem for inclusive gluon production was developed. In
particular, a generalized factorization property of the two-source problem in high energy QCD was derived in which both incoming fluxes are
described by the solutions of the JIMWLK equation (\ref{JIMWLK}) characterized by the corresponding saturation scales. This approach allows to
calculate the inclusive gluon production cross section. Schematically the leading logarithmic correction to such observable ${\cal O}$ can be
written as \cite{HEF1,HEF2,HEF3}:
\begin{equation}
\Delta {\cal O} = \left[ \log(1/x_1) {\cal H}_1 + \log(1/x_2) {\cal H}_2 \right] {\cal O}
\end{equation}
where $\log(1/x_{1,2})$ are the rapidity shifts taken into account in computing the quantum corrections and ${\cal H}_{1,2}$ are the JIMWLK
Hamiltonians of the incident nuclei.

\subsection{Jet quenching and parton energy loss in dense non-abelian medium}

As has been already mentioned in the section 2, one of the most impressive results of the experiments at RHIC is a dramatic decrease in the yield
of particles with large transverse momentum -- a phenomenon directly related to the so-called jet quenching in dense QCD matter. The underlying
phenomenon is, of course, more general than just the above-mentioned attenuation of inclusive spectra of transverse momentum but rather refers to
the specific medium-induced changes in the physics of multiplarticle production. The main focus of jet quenching studies is on looking for
modifications of standard patterns of processes at large transverse momenta (such as jet production in nuclear collisions) as compared to the
results for pp collisions. The jetty structure of final state in events characterized by large transverse energies is one of the generic features
of particle production in high energy hadron collisions. The properties of these jets, i.e. the collimated fluxes of hadrons, are very well
described, especially in the $e^+e^-$-annihilation,  by perturbative QCD under assumption of the local parton-hadron duality. This refers both to
the cross-section and topology of jetty events and the properties of the individual jets such as intrajet particle distributions, multiplicity
distributions, etc. Thus QCD jet physics is a very natural place to look for medium-induced modifications in nuclear collisions -- in the hope
that one can achieve reasonable theoretical and experimental control on its relevant characteristics. For the recent reviews on theoretical
approaches to describing jet quenching and their comparison with experimental data see e.g. \cite{E09,BSZ00,KW03,GVWZ03,Z04a,SS07,W09,M10}.

 Schematically one can break the process of jet formation into three stages:
\begin{itemize}
\item{Production of an initial high energy parton. At asymptotically high energies a relevant theoretical description is given by the
 standard formalism of collinear factorization}
\item{Creation of final state high energy particle is generically accompanied by its radiation. Technically it is convenient to ascribe
 to the initially produced parton some virtuality $Q_0^2$ which is shaken off by subsequent perturbative radiation of gluons and quark-antiquark
 pairs until reaching some cutoff invariant mass $Q^2_h$ at which the perturbative description is no longer applicable.}
\item{Hadronization of partonic configuration created in the course of QCD jet evolution}
\end{itemize}

Let us consider, for example, the simplest inclusive distribution (fragmentation function) $D^h (x=E_h/E_0,Q_0^2)$ describing a probability for a
hadron in the jet originated by the parton\footnote{In this section we will omit all parton subscripts and, for simplicity, consider the purely
gluonic cascade.} with the energy $E_0$ and virtuality $Q_0$ to have the energy $E_h=xE_0$. At sufficiently large $Q_0^2$ one can write
\begin{equation}\label{fragfunvac}
D^h (x,Q_0^2) = \int_x^1 dz \; {\cal P} (z,Q^2_h,Q^2_0) D^h (x,Q_h^2)
\end{equation}
where ${\cal P}_{ab} (z,Q^2_h,Q^2_0)$ is the perturbatively calculable probability of producing a parton with the fractional energy $z=E/E_0$ and
virtuality $Q^2_h$ in the jet originated by the parton with the energy $E_0$ and virtuality $Q_0$ and  $D^h (x=E_h/E,Q_h^2)$ is the
nonperturbative fragmentation function describing the corresponding parton-hadron conversion. Probabilistic description of parton-hadron
transformation is possible due to collinear factorization property of QCD that ensures, at large enough $Q_0^2$,  the absence of interference
between perturbative evolution of the initial parton and the final nonperturbative process of color blanching and formation of the final hadron.
The probability ${\cal P} (z,Q^2_h,Q^2_0)$ refers to a QCD cascade converting the initial parton into a spray of partons with different fractional
energies $\{ z \}$ and the same final virtuality $Q^2_h$. In the in-vacuum QCD the evolution of the fragmentation function $D^h_a(x,Q_0^2)$ with
$Q_0^2$ is described by the standard Dokshitzer-Gribov-Lipatov-Altarelli-Parisi (DGLAP) evolution equations \cite{DGLAP1,DGLAP2,DGLAP3}.

At large $Q_0^2$ the formation of the initial parton takes place at very small time and, in the first approximation, the medium effects on this
initial hard process can be neglected. However, in the in-medium QCD both the probabilistic kernel ${\cal P} (z,Q^2_h,Q^2_0)$ and the
nonperturbative fragmentation function $D^h (x=E_h/E,Q_h^2)$ depend, generally speaking, on the properties of the medium. The dominant opinion in
the current literature is that all the relevant medium-induced effects take place at the perturbative stage and refer to ${\cal P}
(z,Q^2_h,Q^2_0)$ leaving $D^h (x=E_h/E,Q_h^2)$ untouched \footnote{Experimental data which favor it were mentioned in section 2.}. Following this
assumption, one can introduce the medium-modified fragmentation function $D^{h ({\rm med})}_a(x,Q_0^2)$
\begin{equation}\label{fragfunmed}
D^{h ({\rm med})}_a(x,Q_0^2) = \int_x^1 dz \left( {\cal P}_{ab}
(z,Q^2_h,Q^2_0) + \Delta {\cal P}_{ab} (z,Q^2_h,Q^2_0 \vert \; \{ \mu \})
\right)D^h_b(x,Q_h^2),
\end{equation}
where $\Delta {\cal P}_{ab} (z,Q^2_h,Q^2_0 \vert \; \{ \mu \} )$ is the medium-induced change of the in-vacuum probability that depends on the
properties of the medium under consideration that are characterized by the set of parameters $\{ \mu \}$.

To calculate $\Delta {\cal P} (z,Q^2_h,Q^2_0 \vert \; \{ \mu \})$ one has to understand the mechanisms of medium-induced effects such as
medium-induced radiation and elastic energy loss experienced by a parton propagating in the dense non-abelian medium. This is discussed in the
paragraph 3.2.1.

Let us note that both of the two above-described medium-induced mechanisms are well studied in the abelian case in the framework of the physics of
the electromagnetic showers in matter (see e.g. \cite{coscad}). An important ingredient of the physics of electromagnetic showers is accounting for
ionization losses leading to specific distortions of the pattern resulting from medium-induced bremsstrahlung and pair creation \cite{coscad}. The
modifications of QCD cascades taking into account "pionization", i.e. the interaction with additional strongly interacting degrees of freedom,
were analytically studied in \cite{pionizationNAO1,pionizationNAO2,pionizationNAO3,pionizationNAO4} for the case of non-angular ordered cascades
and in  \cite{pionizationAO1,pionizationAO2} for the angular-ordered ones.

\subsubsection{Parton energy loss}

In the previous paragraph we have mentioned that the process of transformation of the initial highly energetic highly virtual parton into the
spray of final hadrons is, in some approximation, that of a perturbative cascading emission of new quanta. Thus we are dealing, e.g., with a
problem of multiple gluon emission. To compute generic in-medium corrections to the in-vacuum picture one has, in principle, to be able to
calculate cross sections for multiple medium-induced gluon emission. As of now this problem is not solved. At the current level of the theory of
the medium-induced radiation one can compute the corresponding cross sections for one-gluon emission in the first order in the strong coupling
constant but, under certain simplifying assumptions, to all orders in the interaction with the medium. The resulting spectrum of one-gluon
emission can be written as a sum of the vacuum and medium-induced contributions
\begin{equation}\label{genspect}
\frac{dI^{\rm tot}}{dw d^2{\bf k}_\perp}=\frac{dI^{\rm vac}}{dw d^2{\bf k}_\perp}
+\frac{dI^{\rm med}}{dw d^2{\bf k}_\perp},
\end{equation}
where
\begin{equation}\label{vacint}
\frac{dI^{\rm vac}}{dw d^2{\bf k}_\perp} = \frac{\alpha_s}{2\pi} \frac{N_c}
{{\bf k}_\perp^2} K^{\rm vac} (z),
\end{equation}
and $K^{\rm vac }(z)=\dfrac{1}{z(1-z)} - 2 + z(1-z)$ is the standard DGLAP gluonic splitting function. The general case with multiple gluon
emission is then considered through constructing a probabilistic branching process obtained by iterating the one-gluon emission described by the
first-order expression (\ref{genspect}). The corresponding theoretical schemes used for constructing such cascades will be discussed in the
paragraph 3.2.2. Below we shall discuss the methods of calculating the second term in the right-hand side of (\ref{genspect}) to the accuracy
$O(\alpha_s)$.

The physical problem at hand is the computation of the leading contribution to the medium-induced energy loss $\Delta E$ of the parton produced
inside the medium on its way from the hot dense fireball. Generically the energy loss $\Delta E$ depends on the medium thickness $L$ and the bulk
characteristics of the medium and is described by some probability distribution ${\cal P} (\Delta E)$.

More precisely, the losses are determined by the following factors:

\begin{itemize}
\item{Probability of an elementary event leading to energy loss  characterized by the opacity $N=L/\lambda$, where $\lambda$ is the mean
free path of the parton in the medium under consideration. For the particle with integrated particle-medium interaction cross section $\sigma$ and
the medium density $\rho$ the mean free path can be estimated as $\lambda \sim 1/(\rho \sigma)$. }
\item{The intensity of the impact or scattering power of the medium characterized by the transport coefficient
${\hat q} = \langle p^2_\perp \rangle / \lambda$, where $\langle p^2_\perp \rangle $ is the average transverse momentum squared that the
propagating particle gets from the elementary act of collision. In the thermal medium ${\hat q}=m^2_D/\lambda$ where $m_D$ is the Debye mass.}
\end{itemize}

There exists a general consensus, supported by extensive model calculations, that the radiative medium-induced energy loss is the dominant one (
see e.g. \cite{Z07}).    Theoretical work on studying the properties of medium-induced gluon radiation was carried out by several groups
\cite{BDMPS1,BDMPS2,BDMPS3,BDMPS4,GLV1,GLV2,GLV3,GLV4,GLV5,GLV6,W00,ASW1,ASW2,ASW3,ASW4,Z96,Z1,Z2,Z99,HT1,HT2,AMY1,AMY2,AMY3,AMY4,AMY5,KP00}. The
approaches of the above-listed groups differ in details of treating the relevant kinematics of the radiation and the details of describing the
medium under consideration. In the present review we will sketch, following \cite{KW03,SS07,W09}, the main steps of the approach developed in
\cite{W00,ASW1,ASW2,ASW3,ASW4,Z96,Z1,Z2,Z99}. The two key ideas allowing to carry out the calculation of the medium-induced radiation in the form
discussed below are the separation of transverse and longitudinal degrees of freedom \cite{Z87} and the usage of the light-cone path integral
formalism \cite{Z96}.

Let us start with the fundamental idea that, at variance with QED, in QCD one can meaningfully consider radiation of gluons in the eikonal
approximation. This radiation takes place due to decoherence of the initial coherent parton flux by soft color rotations caused by the fields of
the medium (see a clear discussion in \cite{KW03,W09}). The basic objects of the eikonal approximation to QCD are the Wilson lines. Staying in the
$A^+=0$ gauge used in the previous section, we get the following expression for the Wilson line in question:
\begin{equation}\label{wl}
W({\bf x}_\perp) = {\cal P} \exp \left[ ig \int dx^+ A^- (x^+,{\bf x}_\perp) \right],
\end{equation}
where ${\bf x}_\perp$ is the coordinate of the incident parton in the impact parameter plane which in the eikonal approximation stays constant and
$A^- (x^+,{\bf x}_\perp)$ are the target fields. The Wilson line (\ref{wl}) fully determines the S-matrix element of the eikonal scattering
\begin{equation}\label{eiksm}
 S(p \to p^{\, \prime}) \sim \delta \left( p^{\; \prime +}-p^+\right) \int d{\bf x}_\perp
 {\rm e}^{-i {\bf x}_\perp ({\bf p}^{\; \prime}_\perp - {\bf p}_\perp )} W({\bf x}_\perp).
\end{equation}
The cross section is therefore fully specified by the average of the product of Wilson lines taken at different points in the impact parameter
plane
\begin{equation}\label{eikcs}
 \left \vert \; S(p \to p^{\, \prime}) \right \vert^2  \;\;\;  \rightarrow \;\;\;
 \frac{1}{N_c} \langle W^\dag ({\bf x}_\perp) W ({\bf y}_\perp) \rangle.
\end{equation}
To elucidate the physical meaning of the average of the product of Wilson lines in (\ref{eikcs}) let us consider the simplest and at the same time
most popular model of the medium in which the medium is described as a collection of static sources of field $a^-({\bf x}_{\perp i})$ located at
points $\{ x^+_i\}$ and thus characterized by the density $n(x^+) = \sum \delta(x^+-x^+_i)$. For such a medium one gets for (\ref{eikcs}):
\begin{equation}\label{eikcs1}
  \frac{1}{N_c} \langle W^\dag ({\bf x}_\perp) W ({\bf y}_\perp) \rangle \backsimeq
  \exp \left \{ -\frac{C_R}{2} \int dx^+ n(x^+) \sigma ({\bf x}_\perp - {\bf y}_\perp) \right \},
\end{equation}
where
\begin{equation}\label{sigmadip}
 \sigma({\bf x}_\perp - {\bf y}_\perp)=\int \frac{d^2 {\bf q}_\perp}{(2 \pi)^2} {\rm Tr}
 \vert \; a^- ({\bf q}_\perp) \vert^2 \left (1 - {\rm e}^{i({\bf x}_\perp - {\bf y}_\perp) {\bf q}_\perp } \right)
\end{equation}
is the cross section of scattering of color dipole on the target under consideration. Let us stress that the dipole cross section appears only in
the matrix element squared (or, equivalently, the cut diagram); the problem at hand is that of scattering a single color probe. In the soft
scattering approximation the dipole scattering cross section is simply expressed as  $ \sigma({\bf x}_\perp - {\bf y}_\perp) \backsimeq C ({\bf
x}_\perp - {\bf y}_\perp)^2$, so that
\begin{equation}\label{eikcssoft}
 \frac{1}{N_c^2-1} \langle W^\dag ({\bf x}_\perp) W ({\bf y}_\perp) \rangle \backsimeq
 \exp \left \{ \frac{1}{4 \sqrt{2}}\int dx^+ {\hat q} (x^+) ({\bf x}_\perp - {\bf y}_\perp)^2 \right \},
\end{equation}
where ${\hat q}(\xi)=2 \sqrt{2}C n(\xi)$. Here ${\hat q}$ is nothing else but the above-introduced scattering power of the medium.

If one is interested in going beyond the soft scattering approximations, it is necessary to take into account the momentum transfer from the
medium. The corresponding elegant generalization of (\ref{eikcs}) reads
\begin{eqnarray}\label{beikcs}
  S(p \to p^{\, \prime}) & \sim & \delta \left( p^{\; \prime +}-p^+\right) p^+ \int d{\bf x}_\perp
  {\rm e}^{-i {\bf x}_\perp) ({\bf p}^{\; \prime}_\perp - {\bf p}_\perp )} \nonumber \\
  & \times & \int {\cal D} {\bf r}_\perp (x^+) \exp \left \{ i \frac{p^+}{2} \int dx^+ \left[ \frac{d  {\bf r}_\perp (x^+)}{dx^+ }\right]\right\}
  W({\bf r}_\perp).
\end{eqnarray}
In (\ref{beikcs}) the momentum transfer from the medium is described by the motion of an effective nonrelativistic particle (a standard situation
in the light-cone quantization) in the impact parameter plane.

The formula (\ref{beikcs}) allows to calculate the spectrum of medium-induced gluon radiation in the medium of longitudinal extent $L^+$ that, in
particular, takes into account the interference of in-vacuum and in-medium contributions \cite{W00} (see also \cite{Z99}):
$$
 k^+ \frac{d I} {d k^+ d^2 {\bf k}_\perp} =  \frac {\alpha_s C_R}{\pi^2 }\frac{1}{{\bf k}_\perp^2} + \frac {\alpha_s C_R}{(2 \pi)^2 k^+} 2 {\rm Re}
 \int_{x^+_0}^{L^+} dx^+_1 \int d^2 {\bf x}_\perp {\rm e}^{-i {\bf k}_\perp {\bf x}_\perp}
$$
$$
 \left [ \frac{1}{k^+} \int_{x^+_2}^{L^+} dx^+_2 {\rm e}^{-\frac{1}{2} \int_{x^+}^{L^+} d \xi n(\xi) \sigma({\bf x}_\perp)}
 \frac{\partial}{\partial {\bf y}_\perp} \cdot   \frac{\partial}{\partial {\bf x}_\perp}
 {\cal K}({\bf x}_\perp, x^+_2 \vert \; {\bf 0},x^+_1) \right. -
$$
\begin{equation}\label{fullspect}
 \left. -2 \frac{{\bf k}_\perp}{{\bf k}_\perp^2} \cdot \frac{\partial}{\partial {\bf y}_\perp}
 {\cal K}({\bf x}_\perp, L^+ \vert \; {\bf 0},x^+_1) \right ].
\end{equation}
Here  ${\cal K} ({\bf r}_\perp (x^+_2) \vert \; {\bf r}_\perp (x^+_1))$ is the Green function
\begin{equation}\label{gf}
 {\cal K} ({\bf r}_\perp (x^+_2) \vert \; {\bf r}_\perp (x^+_1)) =
 \int {\cal D} {\bf r}_\perp \exp \left \{ \int_{x^+_1}^{x^+_2} d \xi \left( i \frac{p^+}{2} {\dot {\bf r}}_\perp^2 -
 \frac{1}{2} n(\xi) \sigma ({\bf r}_\perp) \right) \right \}.
\end{equation}
In the soft scattering approximation the Green function (\ref{gf}) simplifies to
\begin{equation}
 {\cal K} ({\bf r}_\perp (x^+_2) \vert \; {\bf r}_\perp (x^+_1)) =
  \int {\cal D} {\bf r}_\perp \exp \left \{ i \frac{p^+}{2} \int_{x^+_1}^{x^+_2} d \xi
  \left( {\dot {\bf r}_\perp}^2+i\frac{{\hat q}(\xi)}{2 \sqrt{2} p^+}{\bf r}_\perp^2 \right)\right\}.
\end{equation}

The expansion in opacity or, equivalently, in the density of the medium is obtained by expanding (\ref{fullspect}) in the cross-section $\sigma$.

The physics behind the full expression (\ref{fullspect}) can be illustrated by considering a soft gluon emission. The formation time of a gluon is
$\tau_{\rm form} \sim 2\omega/k_\perp^2$. This time should be sufficiently long to ensure the phase change $\Delta \Phi \sim 1$ sufficient for
decoherence and thus radiation. In more details,
\begin{equation}
\Delta \Phi = \langle \frac{L}{\tau_{\rm form}}\rangle \sim \left \langle
\frac{k_\perp^2}{2 \omega} \Delta z \right \rangle = \frac{{\hat q} L}{2 \omega} L
\equiv \frac{\omega_c}{\omega},
\end{equation}
where
\begin{equation}
\omega_c = \frac{1}{2} {\hat q} L^2.
\end{equation}
In the limit $\omega \ll \omega_c$ the medium acts coherently reducing the intensity of radiation while that of $\omega \gg \omega_c$ corresponds
to incoherent scattering.
\begin{eqnarray}
 \left. \omega \frac{dI}{d\omega} \right \vert_{\omega \ll \omega_c} &
 \thickapprox & \alpha_s \sqrt{\frac{{\hat q}L^2}{\omega}}, \nonumber \\
 \left. \omega \frac{dI}{d\omega } \right \vert_{\omega \gg \omega_c} &
 \thickapprox & \alpha_s \frac{{\hat q}L^2}{\omega }.
\end{eqnarray}
In the soft limit $\omega \ll \omega_c$ one has for the average energy loss
\begin{equation}
\langle \Delta E \rangle \backsimeq \int_0^{\omega _c} d\omega \omega
\frac{dI}{d\omega } \sim {\hat q} L^2.
\end{equation}
For some time there existed an opinion that the quadratic dependence on the medium thickness $L$ is a specifically non-Abelian effect. Recently,
however, in \cite{SP09} (see also \cite{Z04}), it was shown that the effect is generic and is valid both in QED and QCD for the radiation of
particle produced inside the medium at the initial piece of its trajectory.

\subsubsection{Multiple gluon emission. In-medium QCD cascades}

At the beginning of this paragraph let us consider the simplest  version of taking into account the multiple medium-induced gluon emission in
which one neglects the evolution of the timelike QCD shower in virtuality. Here, the possibility of multiple emission is taken into account by
constructing a simple Poissonian picture in which the probability distribution of the cumulative energy loss ${\cal P}(\Delta E)$ reads
\begin{eqnarray}\label{pucas}
{\cal P}(\Delta E) & = & p_0 \sum_{n=0}^\infty \frac{1}{n!} \int \left[
\prod_{i=1}^n \int d\omega _i dk_{\perp\;i} \frac{dI^{\rm med}}{d\omega _i dk_{\perp\;i}} \right]
\delta \left(\sum_{i=1}^n \omega _i -E \right), \nonumber \\
 &  & p_0 = \exp \left[ - \int d\omega dk_\perp \frac{dI^{\rm med}}{d\omega dk_\perp} \right].
\end{eqnarray}
The Poissonian cascade described by (\ref{pucas}) can be derived from the full DGLAP evolution (see the Appendix in \cite{ACSX08}). Knowing ${\cal
P}(\Delta E)$ one can compute the distribution ${\cal P} (\varepsilon)$ of the fractional energy  loss $\varepsilon=\Delta E/E_0$ of a parton with
the initial energy $E_0$ and obtain the equation for the medium-modified fragmentation function originally suggested in \cite{WHS96}:
\begin{equation}
{\cal D}^m_{a \to h} (z,Q^2) = \int \frac{d \varepsilon}{1-\varepsilon} \; {\cal P}(\varepsilon) \; {\cal D}_{a \to h} \left(
\frac{z}{1-\varepsilon},Q^2 \right).
\end{equation}

Turning now to the discussion of the complete picture let us stress once again that the energy flow in-vacuum high energy processes involving
quarks and gluons that are characterized by large momentum transfer is dominated by well-collimated jets originating from multiple cascading
radiation of gluons and quark-antiquark pairs. The physical origin of the cascading process is the intense gluon radiation of a high energy parton
created in a hard subprocess. From the operational point of view one can consider this radiation as a cascade degradation of the (large)
virtuality of the initial parton until reaching some non-perturbative scale $Q_h$ at which preconfined colorless clusters and/or QCD strings are
formed.

To make detailed studies of in-vacuum QCD cascades one turns to Monte-Carlo simulations. The two most popular versions of these MC generators are
PYTHIA \cite{PYTHIA} and HERWIG (Hadron Emission Reactions With Interfering Gluons) \cite{HERWIG1,HERWIG2,HERWIG3}. Both of them implement the key
feature of in-vacuum QCD cascades, the angular ordering of successive decays that has its origin in fine-tuned quantum coherence effects. In the
older version of PYTHIA the angular-ordering restrictions were implemented by hand while in its later versions and in HERWIG they are incorporated
by construction by choosing an appropriate evolution variable. In the presence of the medium there appear new mechanisms that change the structure
of in-vacuum QCD jets. In the in-vacuum cascade the only source of evolution is the initial virtuality of a jet. In the in-medium case the medium
acts as a source of "extra" virtuality to the propagating partons resulting in induced radiation and a sink of energy due to the energy loss
resulting from elastic and inelastic collisions of propagating partons with the particles in the medium. The first MC code adding medium-induced
radiation was PYQUEN (Pythia QUENched) \cite{PYQUEN}. Recently there appeared MC codes describing the medium-induced effects in QCD cascades such
as JEWEL \cite{JEWEL}, Q-PYTHIA \cite{QPYTHIA} (both based on the mass-ordered version of PYTHIA) and "Q-HERWIG" \cite{QHERWIG} based on HERWIG.

Let us reiterate that the complete picture combining the in-vacuum and in-medium evolution is still not available. The main difficulty lies in the
fact that the medium created in high energy nuclear collisions is necessarily of some finite extent, so that the modifications induced by the
medium are dependent on its characteristics in space-time: one has to place the cascade inside the medium, see Fig.~1 in which the origin of the
cascade is placed at some distance $L$ from the border separating the dense medium from the normal vacuum.
\begin{figure}[h]\label{fcascade}
 \centering
 \includegraphics[width=0.75\textwidth]{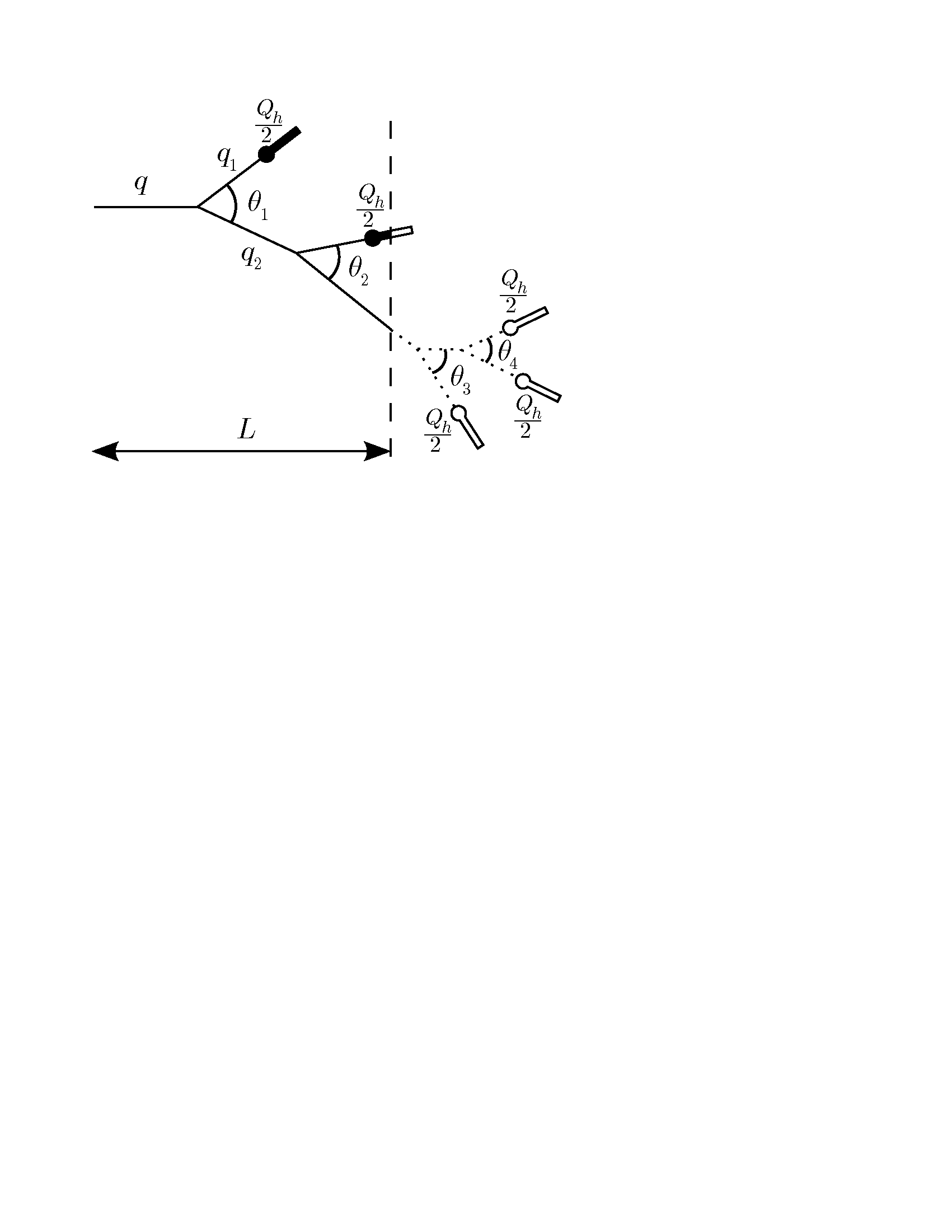}
 \caption{In-medium QCD cascade.}
\end{figure}

Obviously, the larger the distance of parton propagation in the medium, the bigger are the medium-induced losses. The key question is therefore to
choose, for describing the in-medium QCD cascade, the language allowing a credible space-time interpretation. As emphasized in \cite{JEWEL}, the
only scheme allowing such interpretation is the old-style PYTHIA virtuality-ordered cascade where connection to the spatiotemporal pattern is
achieved by calculating the lifetime $\tau$ of a virtual parton which, for a parton with the energy $E$ and virtuality $Q^2$ that has been created
in the decay of its parent parton with the virtuality $Q^2_{\rm par}$, reads
\begin{equation}\label{formtime}
\tau = E \left( \frac{1}{Q^2}-\frac{1}{Q^2_{\rm par}} \right).
\end{equation}
The lifetime of the initial parton is taken to be $\tau_{\rm in}=E_{\rm in}/Q^2_{\rm in}$. For example, the time $\tau_h$ at which the parton with
the momentum $q_1$ in Fig.~1 produced the final prehadron is
$$
\tau_h = E_0 \frac{1}{Q^2} + E_1 \left(
\frac{1}{Q^2_1}-\frac{1}{Q^2}\right) = \frac{E_0}{Q^2} \left[ 1+x
\left( \frac{Q^2}{Q^2_1}-1\right) \right],
$$
where $x=E_1/E_0$. We have already mentioned that in the absolute majority of studies devoted to energy loss in high energy nuclear collisions it
is assumed that hadronization takes place outside of the hot and dense fireball. Setting out the clock with the help of the formula
(\ref{formtime}) helps to address this issue quantitatively \cite{LN10}. For example, for the parton  $q_1$ in Fig.~1 it happened that $\tau_h <
L$ so that both its formation at time $E_0/Q^2_0$ and its decay into final prehadron took place inside the medium. To get a feeling for the
corresponding physics, let us consider, for example, a gluon with the energy $E_0=100 \; {\rm GeV}$ equipped with a typical initial invariant mass
$Q_0=10\;{\rm GeV}$. This gluon  will on the average decay at $\tau_0 \sim E_0/Q_0^2=1\;{\rm GeV}^{-1}\backsimeq 0.2 \; {\rm fm}$, so that if the
vertex in which the initial hard gluon was generated was sufficiently far (several fermies) from the surface of hot fireball, a significant part
of cascade vertices will in fact be generated inside the medium.

In what follows we shall concentrate on the description of the old-style PYTHIA virtuality-ordered cascade and its in-medium modifications. The
cascade evolution goes through a sequence of decays $q \Rightarrow q_1+q_2$, where $q=(E,{\bf q})$ is a four-momentum of the parent gluon and
$q_{1(2)}$ are those of the daughter ones. We study the timelike evolution in which cascading reshuffles the initial jet virtuality in such a way
that at each decay one has $q^2 >q_{1}^2+q_{2}^2$. The cascading process at a given branch stops when a virtuality of the last parton reaches a
threshold level $Q_h^2$, i.e. no decay into two partons with minimal virtuality $Q^2_h/4$ can occur. The key property of the in-vacuum QCD
cascades, effective angular ordering of subsequent decays due to color coherence (i.e. ensuring that, e.g., $\theta_2 < \theta_1$ in Fig.~1), is
not automatically taken into account in the mass-ordered evolution and has to be enforced explicitly by accepting only those generated decays for
which the condition of angular ordering does hold. The basic ingredient of the probabilistic formalism of QCD cascade evolution is the conditional
probability for a parton produced at the scale $Q^2_{\rm in}$ with the energy $E$ to undergo a subsequent decay at the scale $Q^2 < Q^2_{\rm in}$
into two daughter partons with the energies $zE$ and $(1-z)E$ correspondingly that can be calculated from the Sudakov formfactor
\begin{equation}\label{sff}
S(Q^2 \vert \; E, Q^2_{\rm in};Q_h^2)= \\ \exp \left[
-\int_{Q^2}^{Q_{\rm in}^2} d{\bf k}_\perp^2 \int_{z_{-}(E,\;{\bf k}_\perp^2
\vert \; Q_h^2)}^{z_{+}(E,\;{\bf k}_\perp^2 \vert \;  Q_h^2)} dz  \frac{dI^{\rm tot}}{dz d^2{\bf k}_\perp} ({\bf k}_\perp^2,z) \right],
\end{equation}
where $z_{\pm}(E,\;k_\perp^2 \vert \;  Q_h^2)$ define the kinematic limits available for the parton decay
\begin{equation}\label{zpm}
 z_{\pm}(E,Q^2 \vert \;  Q_h^2)
 =\frac{1}{2} \left(1 \pm \sqrt{\left(1-\frac{Q^2}{E^2}\right) \left(1-\frac{Q_h^2}{Q^2}\right)} \right),
\end{equation}
and where the full differential probability of gluon emission $dI^{\rm tot}/dz d^2{\bf k}_\perp$ was defined in (\ref{genspect}).

When considering the properties of the in-medium QCD cascades one has to deal with several mechanisms modifying the in-vacuum intrajet
characteristics, in particular the loss of quantum coherence and the resulting disruption of angular ordering for decays inside the medium, the
medium-induced radiation (inelastic energy loss) and elastic scattering of partons on particles in the medium (elastic energy loss). Below we
shall describe the corresponding modifications of the MC procedure for generation of the in-vacuum QCD cascades that take these features into
account.

Let us start with the decoherence-induced disruption of angular ordering \cite{LN10}. The violent environment created in ultrarelativistic heavy
ion collisions acts as a source of random energy-momentum and color with respect to the diagrams for the in-vacuum processes. Generally speaking,
this kind of random impact ruins the phase tuning lying at the heart of quantum interference phenomena. The situations in which random external
influences destroy the interference-related effects are very common in solid state physics. A well-studied example is provided by the destruction
of the weak localization in the presence of random external impacts, see e.g. \cite{B84,CS86}. Thus it is intuitively quite clear that the quantum
coherence effects  should be broken in the violent environment created in heavy ion collisions. This was first explicitly mentioned in
\cite{JEWEL}, in which the origin of decoherence and ensuing disappearance of angular ordering was related to rescattering of cascading particles
on scattering centers in the medium. Another argument in support of the assumption of angular decoherence is the fact that, at least at the level
of Vlasov approximation, the characteristic time of color randomization is much less than the characteristic scattering time \cite{SG1,SG2}. As
color randomization alone is sufficient for angular decoherence, one could expect that this effect is even stronger than assumed in \cite{JEWEL}.
The simplest assumption taking into account this effect is that angular ordering is broken for all the decays that take place inside the medium.
The effect is thus directly dependent on size of the medium available for the cascade development. In terms of notations introduced in Fig.~1, the
effect is $L$-dependent. The corresponding modifications of the  rapidity distributions are shown in Fig.~2. We see that the decoherence effects
alone, i.e. before considering medium-induced energy loss, lead to substantial softening of the intrajet particle spectrum.

\begin{figure}[h]
 \centering
 \includegraphics[width=0.75\textwidth,height=0.5\textheight]{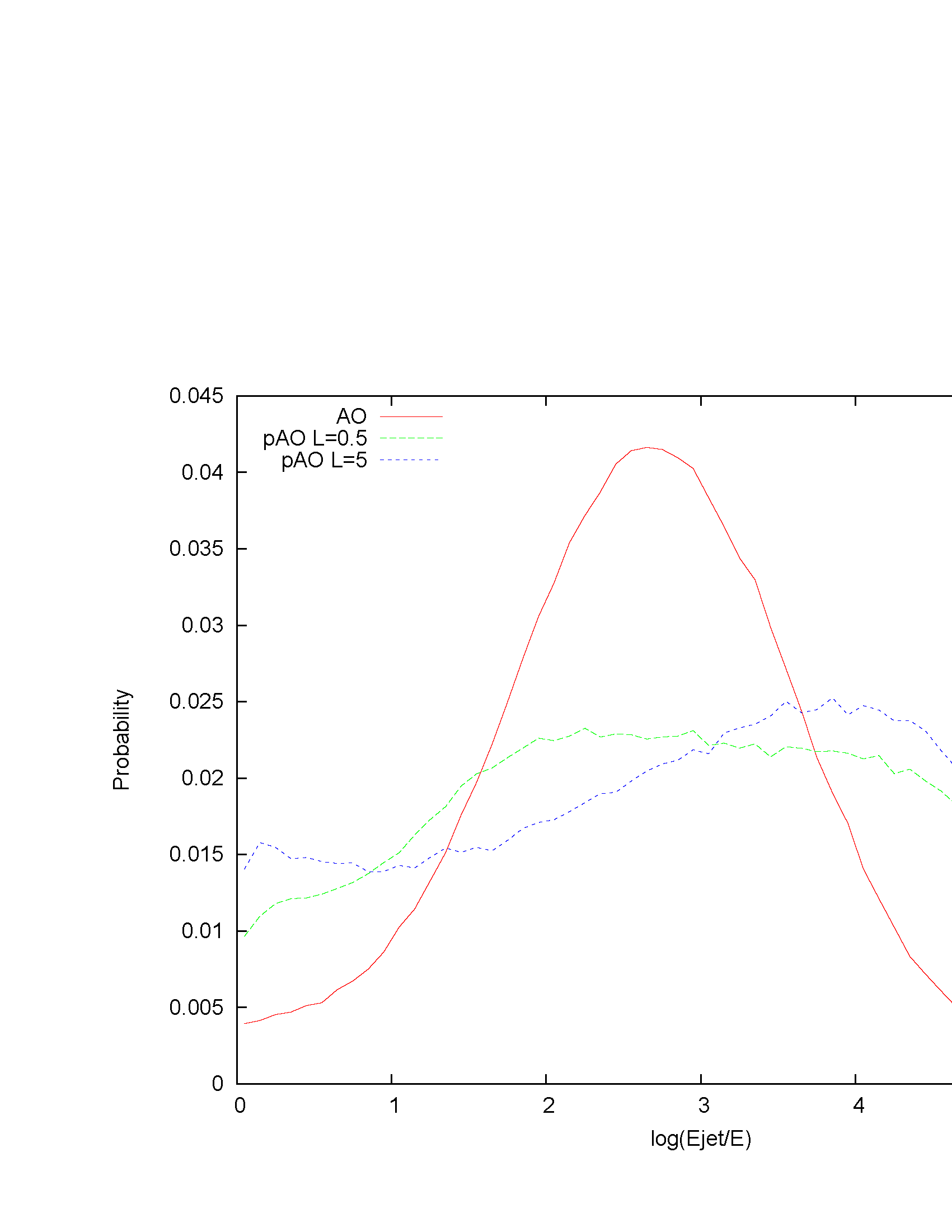}
 \caption{Distribution in rapidity $P(y)$ of final prehadrons: 1) $L=0\,{\rm fm}$, full angular ordering, red, full line; 2) $L=0.5\,{\rm fm}$,
 partial angular ordering, green, dashed line; 3) $L=5\,{\rm fm}$, partial angular ordering, blue, dotted line \cite{LN10}.}
\end{figure}\label{paonl}

It is important to notice that, as seen from Fig.~2, the softening of the  intrajet rapidity distribution sets in at relatively small medium sizes
$L \sim 1\;{\rm fm}$.

Let us now turn to the effects of the medium-induced radiation. A rigorous treatment of this problem is very difficult. Only very recently in
\cite{MCLPM1,MCLPM2} an elegant formalism allowing to take into account LPM effect in the MC cascades was described.

The results of modifications of QCD cascade properties were obtained using simpler schemes based on various approximations for the effective
differential spectrum (\ref{genspect}) or, equivalently, to the modified splitting function
\begin{equation}
K^{\rm tot}(z,E,Q^2,{\hat q},L)=K^{\rm vac}(z) + \Delta K (z,Q^2,{\hat q},L)
\end{equation}
that depends, generally speaking, not only on the energy ratio $z$ but also on the energy and virtuality of the parent parton, stopping power of
the medium, path length in the medium, etc. \cite{WG01}.

The simplest version of the medium-modified splitting function mimicking the effects of medium-induced radiation was suggested in \cite{BW05}:
\begin{equation}\label{splitfmed}
K_{\rm vac}(z) \to (1+f_{\rm med}) K_{\rm vac}(z)
\end{equation}
A more advanced definition of the medium-induced splitting function was proposed \cite{ACSX08}:
\begin{equation}\label{splitspectrum}
\Delta K(z,E,Q^2,{\hat q},L) \backsimeq \frac{2 \pi Q^2}{\alpha_s} \frac{dI^{\rm med}}{dz dQ^2}
\end{equation}
where $dI^{\rm med}/dz dQ^2$ is taken from Eq.~(\ref{fullspect}).

With the concrete expressions for the modified splitting functions like (\ref{splitfmed}),(\ref{splitspectrum}) the main ingredient for the
description of the cascade, the Sudakov formfactor (\ref{sff}), is fully specified.

The main quantity addressed in these MC simulations is the jet quenching ratio $R_{AA}(p_\perp)$. The agreement with the experimental data is
achieved by tuning the parameters of the medium like ${\hat q}^2$ or $f^{\rm med}$. In most cases the agreement can indeed be reached - however,
at the expense of very large variations in the values of these parameters obtained within different formalisms and of the numerous theoretical
uncertainties.

As to the more generic intrajet characteristics, the general expectation concerning the effect of the medium-induced interactions are:
\begin{itemize}
\item{Softening of the rapidity distributions.}
\item{Broadening of the distributions in transverse momentum.}
\item{Increased multiplicity.}
\end{itemize}
These expectations are indeed confirmed by the studies undertaken with the above-described MC generators \cite{JEWEL,QPYTHIA}, see also \cite{LMPSAT09}.
More involved properties of medium-modified cascades such as higher order multiplicity moments were also studied in the literature \cite{DS06,APQ09}.

\section{The macroscopic approach to the quark-gluon medium}

Macroscopic collective properties of a medium may reveal itself in its mechanical motion as a whole (e.g., viscosity) described by hydrodynamics
or in its response to external color currents (e.g., chromopermittivity) described by in-medium QCD. Here we mostly deal with the latter approach
discussing hydrodynamics at the end very briefly because there exist many extensive reviews of it (see, e.g., \cite{sh1, hein}).

Considering the evolution of energy flows due to the elastic scattering, bremsstrahlung or synchrotron radiation one follows the fate of an
initial source (a particle, a parton, a dipole etc) and the medium impact on these processes. The evolution of its velocity vector is crucial for
conclusions about its energy losses. The notion of the coherence length becomes important when multiple rescattering is taken into account.

However, the medium itself can radiate in response to permutations. The genuine role of the medium and its collective properties are most clearly
revealed by its polarization ${\bf P}$ due to the external current. The macroscopic approach to the description of such collective properties is
the most suitable one. The linear response of the medium to the electromagnetic field ${\bf E}$ is usually described as
\begin{equation}
{\bf P}= \frac {\epsilon - 1}{4\pi }{\bf E},     \label{pe}
\end{equation}
where $\epsilon $ is the dielectric permittivity. It  is seen that the polarization can be quite strong for large values of $\epsilon $. The
initial radiation process serves as a trigger for the collective response of the medium initiated by the polarization. The well known examples are
provided by the Cherenkov radiation, the wake and the transition radiation.

What is typical for their description is  the (approximate) constancy of the particle velocity vector used in the external current. In
ultrarelativistic processes $(\gamma \gg 1$) the relative change of the velocity is much smaller than the relative energy loss because they are
connected by the formula
\begin{equation}
\frac {\Delta E}{E}=(\gamma ^2-1)\frac {\Delta v}{v}  \label{ev}
\end{equation}
and the above statement is well supported. The velocity loss can become noticeable only for non-relativistic partons.

In what follows we consider very high energy processes. It is well known that the gluons become the main component of the wave functions of the
colliding hadrons. At LHC, the $gg$-luminosity is at least by an order of magnitude higher than the $\sum qq$-luminosity. Therefore the in-medium
gluodynamics \cite{inmed} is considered below. It simplifies the formulae. Quarks can be easily included too \cite{dpri}.

\subsection{Equations of in-medium QCD}

The in-medium equations of gluodynamics differ from the in-vacuum equations by taking into account the chromopermittivity of the quark-gluon
medium. Similar to the dielectric permittivity in electrodynamics it describes the linear response of the matter to passing partons. At the
leading order in the coupling constant the equations are completely analogous to electrodynamical ones with chromopermittivity just replacing the
dielectric permittivity.

The classical in-vacuum Yang-Mills equations are
\begin{equation}
\label{f.1}
D_{\mu}F^{\mu \nu }=J^{\nu },
\end{equation}
\begin{equation}
\label{1}
F^{\mu \nu }=\partial ^{\mu }A^{\nu }-\partial ^{\nu }A^{\mu }-
ig[A^{\mu },A^{\nu }],
\end{equation}
where $A^{\mu}=A_a^{\mu}T_a; \; A_a (A_a^0\equiv \Phi_a, {\bf A}_a)$ are the gauge field (scalar and vector) potentials, the color matrices $T_a$
satisfy the relation $[T_a, T_b]=if_{abc}T_c$, $\; D_{\mu }=\partial _{\mu }-ig[A_{\mu }, \cdot], \;\; J^{\nu }(\rho, {\bf j})$ is a classical
source current, and the metric tensor is $g^{\mu \nu }$=diag(+,--,--,--).

In the covariant gauge $\partial _{\mu }A^{\mu }=0$ they are written as
\begin{equation}
\label{f.2}
\square A^{\mu }=J^{\mu }+ig[A_{\nu }, \partial ^{\nu }A^{\mu }+F^{\mu \nu }],
\end{equation}
where $\square $ is the d'Alembertian operator. Thus the leading term in the expansion in the coupling constant for the classical gluon  field is
given by the solution of the corresponding abelian problem.

The chromoelectric and chromomagnetic fields are
\begin{equation}
\label{2}
E^{\mu}=F^{\mu 0 },
\end{equation}
\begin{equation}
\label{3}
B^{\mu}=-\frac {1}{2}\epsilon ^{\mu ij}F^{ij},
\end{equation}
or, as functions of the gauge potentials in vector notation,
\begin{equation}
\label{4}
{\bf E}_a=-{\rm grad }\Phi  _a-\frac {\partial {\bf A}_a}{\partial t}+
gf_{abc}{\bf A}_b \Phi _c,
\end{equation}
\begin{equation}
\label{5}
{\bf B}_a={\rm curl }{\bf A}_a-\frac {1}{2}gf_{abc}[{\bf A}_b{\bf A}_c].
\end{equation}
The equations of motion (\ref{f.1}) in vector form are written as
\begin{equation}
\label{6}
{\rm div } {\bf E}_a -gf_{abc}{\bf A}_b {\bf E}_c = \rho _a,
\end{equation}
\begin{equation}
\label{7}
{\rm curl } {\bf B}_a-\frac {\partial {\bf E}_a}{\partial t} - gf_{abc}
(\Phi _b {\bf E}_c+[{\bf A}_b {\bf B}_c])= {\bf j}_a.
\end{equation}

The effects due to the medium are accounted for if $\rho _a$ and ${\bf j}_a$ contain both external and internal (induced) terms. However,
analogously to electrodynamics, the role of internal currents may be represented in the linear response approach by the medium permittivity
$\epsilon $ if $\bf E$ is replaced by ${\bf D} =\epsilon {\bf E}$ in $F^{\mu \nu} \;$, i.e. in Eq. (\ref{2}). After that is done, only external
currents should be considered in the right hand sides of the equations. Therefore Eqs (\ref{6}), (\ref{7}) in vector form are most suitable for
generalization to the in-medium case.

In terms of potentials the equations of {\it in-medium} gluodynamics are cast
in the form \cite{inmed}
\begin{eqnarray}
\bigtriangleup {\bf A}_a-\epsilon \frac{\partial ^2{\bf A}_a}{\partial t^2}=
-{\bf j}_a -
gf_{abc}(\frac {1}{2} {\rm curl } [{\bf A}_b, {\bf A}_c]+
\epsilon \frac {\partial }
{\partial t}({\bf A}_b\Phi _c)+\frac {1}{2}[{\bf A}_b {\rm curl } {\bf A}_c]-  \nonumber \\
\epsilon \Phi _b\frac
{\partial {\bf A}_c}{\partial t}-
\epsilon \Phi _b {\rm grad } \Phi _c-\frac {1}{2} gf_{cmn}
[{\bf A}_b[{\bf A}_m{\bf A}_n]]+g\epsilon f_{cmn}\Phi _b{\bf A}_m\Phi _n),
\hfill \label{f.6}
\end{eqnarray}

\begin{eqnarray}
\bigtriangleup \Phi _a-\epsilon \frac {\partial ^2 \Phi _a}
{\partial t^2}=-\frac {\rho _a}{\epsilon }+
gf_{abc}(-2{\bf A}_c {\rm grad }\Phi _b+{\bf A}_b
\frac {\partial {\bf A}_c}{\partial t}-\epsilon
\frac {\partial \Phi _b}{\partial t}
\Phi _c)+  \nonumber  \\
g^2 f_{amn} f_{nlb} {\bf A}_m {\bf A}_l \Phi _b. \hfill  \label{f.7}
\end{eqnarray}
If one neglects the terms with  explicitly shown charge $g$, one gets the set of abelian equations, which differ from electrodynamical equations
by the color index $a$ only. The most important property of the solutions of these equations is that while the in-vacuum ($\epsilon = 1$)
equations do not admit any radiation processes, for $\epsilon \neq 1$ there appear solutions of these equations with non-zero Poynting vector even
in the classical approach. They predict such coherent effects as Cherenkov gluons, the wake and transition radiations (see section 4.3). This
corresponds well to the microscopic description in which the matter, at early times after a heavy ion collision, is described in terms of the
coherent classical field.

As pointed above, ${\bf j}_a$ is treated now as an external current ascribed to partons moving fast relative to the other partons "at rest". It is
proportional to $g$ as seen from Eq. (\ref{f.11}) below. The potentials are also of the order of $g$ if one neglects the $g$-dependence of
$\epsilon $ discussed in the section 4.2. Thus the terms with explicitly shown $g$ in the right hand sides of Eqs (\ref{f.6}), (\ref{f.7}) would
be of the order $g^3$. The higher order corrections may be calculated (see section 4.5) and quite interesting conclusion about the color rainbow
is obtained.

\subsection{The chromopermittivity}

The most economical way of description of matter properties is to take into account an impact of internal currents in the medium
phenomenologically by the chromopermittivity $\epsilon $ introduced as a constant in Eqs (\ref{f.6}), (\ref{f.7}). Its spatio-temporal dependence
may be included similar to electrodynamics. For the Fourier components of $\epsilon $ this would imply the dependence on the frequency $\omega $
and the wave vector $\bf k$. It is well known in electrodynamics that the magnetic permeability is then automatically taken into account.
Therefore it is necessary to consider the permittivity tensor and its color dependence.

These complications can become important but even in electrodynamics they are often ignored. The medium is often assumed to be isotropic and
homogeneous. Methods of linear response theory are applied. The $\omega $-dependence is commonly accounted for because the dielectric permittivity
of ordinary substances depends on $\omega $. Moreover, it has an imaginary part determining the absorption. The energy behavior of $\epsilon $
determines the scales specific for a particular substance.

All these features are usually found experimentally using  the relation between the dielectric permittivity and the refraction index $n$
\begin{equation}
\label{f.18}
\epsilon =n^2
\end{equation}
For example, both ${\rm Re }\, \epsilon $ and ${\rm Im }\, \epsilon $ for water have been measured in a wide energy interval ranging over 20
orders of magnitude \cite{ja}. The real part ${\rm Re }\, \epsilon $ is approximately constant in the visible light region ($\sqrt {\epsilon
}\approx 1.34$), increases at low $\omega $ (up to $\epsilon \approx 80$) and becomes smaller than 1 at high energies, tending to 1 at frequencies
exceeding the Langmuir (plasma) frequency $\omega _L$ as
\begin{equation}
{\rm Re }\epsilon = 1-\frac {\omega _L^2}{\omega^2}.         \label{eed}
\end{equation}
The imaginary part (${\rm Im }\, \epsilon $) responsible for the absorption is very small for visible light but drastically increases in nearby
regions both at low and high frequencies (in this way the Nature saves our eyes!).

Thus there exist at least two characteristic scales for ${\rm Re } \epsilon $: just below the optical region and the high energy scale determined
by the Langmuir frequency related to the proper plasma oscillations. The borders of the optical region define the scales for ${\rm Im }\epsilon $.
Theoretically this behavior is ascribed to various collective excitations in water relevant to its response to radiation with different
frequencies. Among them the resonance excitations are quite prominent (see, e.g., \cite{fe}). Even in electrodynamics, the quantitative theory of
this behavior is still lacking, however. Moreover the formula (\ref{eed}) is purely electrodynamical one and does not take into account hadronic
processes at extremely high energies where the hadronic components of photon wave functions are important.

Then, what can we say about the chromopermittivity?

The chromopermittivity can be calculated from the polarization operator. Up to now, the attempts to calculate the chromopermittivity from first
principles are not very convincing. The corresponding dispersion branches have been computed in the lowest order perturbation theory \cite{kk,
kl1, kl2, we}. The properties of collective excitations have been studied in the framework of the thermal field theories \cite{bi, rrs1, rrs2,
amy, qmsh}. In these works the simplest perturbative results representing $\epsilon $ are generalized by summing some types of diagrams. Only the
time-like dispersion relations were obtained in this way. The space-like dispersion relation was obtained in \cite{ko} in the framework of the
method using an {\it ad hoc} assumption about a special role of scalar resonances. Some studies were performed in \cite{dpri}. All of these works
refer to the weakly coupled medium. The phenomenological model built upon some intuitive guesses (see, e.g., \cite{rup}) were also studied. The
physical consequences of introducing the chromopermittivity could be mimiced in microscopic approach considering different non-zero masses and
coupling constants of in-vacuum and in-medium gluons (see, e.g., \cite{Z96, Z1, Z2, zakh, cs, dgyu1, dgyu2, djor}). That allows to vary the
dispersion relations. However, no convincing principles of their choice exist and the number of adjustable energy dependent parameters increases
even though they are related with more fundamental microscopic notions.

Experimental data surely favor strong response of the medium to deposited energy. It implies that the chromopermittivity must be defined in terms
of non-perturbative in-medium gluon field correlation functions that are as yet unknown. Therefore, it makes sense to consider the
chromopermittivity as a complicated function of $g$ and, correspondingly, the Eqs (\ref{f.6}), (\ref{f.7}) as strongly non-linear in $g$ even at
the classical level because $\epsilon \neq 1$ takes into account some quantum and non-abelian effects.

In view of this situation, we prefer to use the general formulae of the scattering theory \cite{ja, go} to estimate the chromopermittivity. It can
be expressed \cite{ja, go, scad} through the real part of the forward scattering amplitude ${\rm Re}F_0(\omega)$ for the corresponding
quantum\footnote{In electrodynamics these quanta are photons, in QCD they are gluons.}
\begin{equation}
\label{f.19}
{\rm Re} \Delta \epsilon = {\rm Re} \epsilon (\omega ) -1= \frac {4\pi N_s
{\rm Re} F_0(\omega)}{\omega ^2}=\frac {N_s\sigma (\omega )\rho (\omega )}
{\omega }
\end{equation}
with
\begin{equation}
{\rm Im} F_0(\omega )=\frac {\omega }{4\pi }\sigma (\omega ).
\end{equation}
Here $\omega$ denotes the energy, $N_s $ the density of scattering centers, $\sigma (\omega)$ the cross section and $\rho (\omega)$ the ratio of
real to imaginary parts of the forward scattering amplitude $F_0(\omega)$. The sign of $\Delta \epsilon $ coincides with the sign of $\rho (\omega
)$.

Unfortunately, we are unable to calculate directly in QCD these characteristics of gluons and have to rely on analogies and our knowledge of the
properties of hadron interactions. The experimental facts we get for this medium are brought about only by particles registered at the final
stage. We hope that some of their features are also relevant for gluons as the carriers of strong interactions. Those are the resonant behavior of
amplitudes at rather low energies and the positive real part of the forward scattering amplitudes at very high energies for hadron-hadron ($pp,
Kp, \pi p$) and photon-hadron ($\gamma p$) processes as measured from the interference of the Coulomb and hadronic parts of the amplitudes. The
latter feature is directly connected with the specific property of hadronic processes, the energy increase of the total cross sections, by the
dispersion relations for forward scattering amplitudes.

Any Breit-Wigner amplitude has a positive (negative) real part below (above) its peak. The combined effect of all collective resonance excitations
in the medium may lead to positive $\Delta \epsilon $ in a wide energy interval at low energies. This is the necessary condition for Cherenkov
effects. As explained in the paragraph 4.3.1, the double-humped events at RHIC and the asymmetry of mass shapes of resonances passing through the
quark-gluon medium observed at SPS may be explained in this way.

At very high energies above the threshold $\omega _{th}$ of the positiveness of real parts of the amplitudes in (\ref{f.19}) one could try to use
a simple model with
\begin{equation}
{\rm Re }\epsilon = (1+\frac {\omega _0^2}{\omega ^2})
\Theta (\omega -\omega _{th}).   \label{eom}
\end{equation}
The model (\ref{eom}) satisfies the Kramers-Kronig requirement for ${\rm Re }\epsilon $ to be an even function and approaches 1 at infinity. If
experimental data on ${\rm Re } F_0(\omega )$ for hadrons are inserted in Eq. (\ref{f.19}) one gets the shape of ${\rm Re }\epsilon $ very close
to Eq. (\ref{eom}) (see Fig. 1 in \cite{d1, d0}). The cosmic ray events with the ringlike (conical) structure may be then explained.

At intermediate energies in between these two intervals the similar arguments would favor negative values of $\Delta \epsilon $. Thus, no
Cherenkov gluons would be emitted with such energies. The transition  gluon radiation (section 4.3.3) is not forbidden, however. The wake effects
(section 4.3.2) can be shown to be small (see the paragraph 4.3.2).

It is tempting to conclude that energy scales determining the behavior of the real part of the chromopermittivity are not directly related either
to the temperature $T$ or to the QCD scale $\Lambda _{QCD}$ of the order of hundreds MeV but extend to higher values defined by the end of the
resonance region (about several GeV) and by the threshold value $\omega _{th}$  connected with energy increase of hadronic total cross sections
(about tens of GeV). The microscopic nature of their origin is yet unclear.

There is another important problem of the definition of the chromopermittivity. The in-medium equations (\ref{f.6}), (\ref{f.7}) are not
Lorentz-covariant. In macroscopic electrodynamics the rest system of the matter is well defined and the dielectric permittivity is considered just
there. For collisions of two nuclei (or hadrons) the definition of $\epsilon $ depends on the experiment geometry. Therefore one needs to consider
the Lorentz-covariant tensor of the chromopermittivity. The spectra of classical Cherenkov radiation in moving media were calculated in \cite{bol}
and with quantum corrections in \cite{alf}.

The notions of dilute and dense systems (i.e. the value of $N_s$ in
(\ref{f.19})) depend on values of $x$ and $p_T$ at which the wave functions
of colliding nuclei are probed.
Due to Lorentz transformations, partons moving in different
directions with different energies can "feel" different states of matter in
the ${\it same}$ collision of two nuclei because of the dispersive dependence
of the chromopermittivity. We discuss this when dealing with different
experimental data and with the problem of instabilities of the
quark-gluon medium (section 4.4).

\subsection{Classical polarization effects in the quark-gluon medium and its
chromodynamical properties}

In view of the similarity between classical QCD and QED equations it would be of no surprise to observe some effects in nucleus collisions
reminding those in electrodynamics. Actually, this idea was first promoted in \cite{d1, d0} relying on the similarity of quarks to electrons and
gluons to photons. The emission of Cherenkov gluons analogous to Cherenkov photons was predicted and used for interpretation of a cosmic ray event
just observed \cite{apan}. These events are discussed in the paragraph 4.3.1.b. The strong support this idea got, however, much later from RHIC
data on nucleus-nucleus collisions (see the paragraph 4.3.1.a).

Before delving in experimental data let us describe the classical solutions of Eqs (\ref{f.6}), (\ref{f.7}). As usually, the current with constant
velocity ${\bf v}$ along the $z$-axis is considered:
\begin{equation}
\label{f.11}
{\bf j}({\bf r},t)={\bf v}\rho ({\bf r},t)=4\pi g{\bf v}
\delta({\bf r}-{\bf v}t).
\end{equation}
The color index is omitted because it points in a fixed direction in color group space. The classical lowest order solution of in-medium
gluodynamics can be cast in the form \cite{inmed, kruk}
\begin{equation}
\label{f.12}
\Phi ^{(1)}({\bf r},t)=\frac {2g}{\epsilon }\frac {\Theta
(vt-z-r_{\perp }\sqrt {\epsilon v^2-1})}{\sqrt {(vt-z)^2-r_{\perp} ^2
(\epsilon v^2-1)}},
\end{equation}
and
\begin{equation}
{\bf A}^{(1)}({\bf r},t)=\epsilon {\bf v} \Phi ^{(1)}({\bf r},t),   \hfill  \label{f.13}
\end{equation}
where the superscript (1) indicates the solutions of order $O(g)$ (not to count for $g$-dependence of $\epsilon $), $r_{\perp }=\sqrt {x^2+y^2}$
is the cylindrical coordinate; $z$ is the symmetry axis. These formulae describe both the space-time profile of the energy-momentum deposition and
the dynamical response of the medium (the value of $\epsilon $).

This solution describes the cone-like emission of Cherenkov gluons at the typical angle
\begin{equation}
\label{f.10}
\cos \theta = \frac {1}{v\sqrt {\epsilon }}.
\end{equation}
It is constant for constant $\epsilon >1$.

The expression for the intensity of the radiation is given by the Tamm-Frank formula \cite{tfra} (up to the Casimir factors $C_R$)
\begin{equation}
\label{f.17}
\frac {dE}{dl}=4\pi \alpha_S C_R\int \omega d\omega (1-\frac {1}{v^2\epsilon
(\omega )})\Theta (v^2\epsilon (\omega )-1).
\end{equation}
For absorbing media $\epsilon $ acquires the imaginary part. The sharp front edge of the shock wave (\ref{f.12}) is smoothed.  The angular
distribution of Cherenkov radiation widens. The $\delta $-function at the angle (\ref{f.10}) is replaced by the a'la Breit-Wigner angular shape
\cite {gr, dklv, wake} with the maximum at the same angle (for small imaginary part of $\epsilon $) and the width proportional to the imaginary
part (see Eq. (\ref{9}) below). The special role of $1/\epsilon $ term in (\ref{f.12}) which describes the wake left behind the parton is
discussed in section 4.3.2.

Let us stress that this classical effect has collective non-perturbative origin even though $\alpha _S$ enters seemingly linearly in Eq.
(\ref{f.17}). The chromopermittivity $\epsilon $ takes into account the non-perturbative terms responsible for the collective medium response.
Only the emission of a primary gluon which triggers this response is treated perturbatively, hence the factor of $\alpha _S$ in Eq. (\ref{f.17}),
but its effect is strongly enhanced and modified by the medium as seen from the term in brackets in the integral.

\subsubsection{Cherenkov gluons}
\medskip

{\bf a. The double-humped structure at RHIC}

The scalar and vector potentials in the momentum space can be cast in the form
\begin{equation}
\Phi_a^{(1)}=2\pi gQ_a\frac {\delta (\omega -kv\zeta)v^2\zeta ^2}{\omega ^2
\epsilon(\epsilon v^2\zeta ^2-1)},
\label{phi}
\end{equation}

\begin{equation}
A_{z,a}^{(1)}=\epsilon v\Phi_a^{(1)},
\label{a}
\end{equation}
\begin{equation}
\zeta=\cos \theta ,
\end{equation}
$\omega , k $ are the energy and momentum, $\theta $ is the polar angle. Let us note again that both the Cherenkov term $(\epsilon v^2\zeta
^2-1)^{-1}$  and the wake factor $\epsilon ^{-1}$ are present.

The energy loss $dW$ per the length $dz$ is determined by the formula
\begin{equation}
\frac {dW}{dz}=-gE_z.    \label{eloss}
\end{equation}
In the lowest order
\begin{equation}
E_z^{(1)}=i\int \frac {d^4k}{(2\pi )^4}[\omega A_z^{(1)}({\bf k},\omega)-
k_z\Phi^{(1)}({\bf k},\omega)]e^{i({\bf k}{\bf v}-\omega)t}.
\label{ez}
\end{equation}
Inserting (\ref{phi}), (\ref{a}) in (\ref{ez}), (\ref{eloss}) one gets
(see also \cite{tgyu})
\begin{equation}
\frac {dW^{(1)}}{dzd\zeta d\omega }=\frac {g^2\omega }{2\pi ^2v^2\zeta }
{\rm Im}\left(\frac {v^2(1-\zeta ^2)}{1-\epsilon _t v^2\zeta ^2}-
\frac {1}{\epsilon _l }\right),
\label{dwdz}
\end{equation}
The first term in the brackets corresponds to the transverse gluon Cherenkov radiation (index $t$ at $\epsilon _t$) and the second term to the
radiation due to the longitudinal wake (index $l$ at $\epsilon _l$). The transverse and longitudinal components of the chromopermittivity  tensor
are explicitly indicated here even though they are equal in any homogeneous medium.

Similar to electrodynamics \cite{gr}, one easily gets from (\ref{dwdz}) the energy-angular spectrum of emitted gluons \cite{dklv, wake} per the
unit length
\begin{equation}
\frac {dN^{(1)}}{d\Omega d\omega }=\frac {\alpha _SC\sqrt x}{2\pi ^2}\left [
\frac {(1-x)\Gamma _t}{(x-x_0)^2+(\Gamma _t)^2/4}+\frac {\Gamma _l}{x}\right ],
\label{9}
\end{equation}
where
\begin{equation}
x=\zeta ^2, \;\;\;
x_0=\epsilon_{1t}/\vert \epsilon _t \vert ^2v^2, \;\;\;
\Gamma _j=2\epsilon_{2j}/\vert \epsilon _j \vert ^2v^2, \;\;\;
\epsilon _j=\epsilon _{1j}+i\epsilon _{2j}.   \label{xep}
\end{equation}
The real ($\epsilon_1$) and imaginary ($\epsilon_2$) parts of $\epsilon $ are taken into account. The angle $\theta $ is the polar angle if the
away-side jet axis would be chosen as $z$-axis. It is clearly seen from Eq. (\ref{9}) that the transverse and longitudinal parts of the
chromopermittivity are responsible for the distinctly different effects. The ringlike Cherenkov structure (conical emission) around this axis is
clearly exhibited in the first term of (\ref{9}). The second term defined by the longitudinal part of $\epsilon $ is in charge of the wake
radiation described in section 4.3.2.

Let us consider the first term. The angle $\theta $ is related to the laboratory polar ($\theta _L$) and azimuthal ($\phi _L$) angles in RHIC
experiments through
\begin{equation}
x=\cos ^2 \theta = \sin ^2 \theta _L \cos ^2\phi _L . \label{tpl}
\end{equation}
Integrating (\ref{9}) over $\theta _L$ one gets \cite{dklv} the final formula to compare with the two-hump structure of azimuthal ($\phi _L$)
correlations observed at RHIC. The formula is quite lengthy and therefore not reproduced here. It is seen already from (\ref{9}) and (\ref{tpl})
that this projection of the two-dimensional ring on its diameter is symmetrical about $\phi _L=\pi $ and exhibits humps whose positions are mostly
determined by $\epsilon _1$ and widths determined by $\epsilon _2$. The first term in (\ref{9}) clearly displays the  Breit-Wigner angular hump
which replaces the $\delta $-functional angular dependence characteristic for real $\epsilon $ (easily obtained from (\ref{9}) at $\Gamma _t
\rightarrow 0$).

Actually, it is a matter of experiment to demonstrate whether Cherenkov gluons can be observed in some energy regions. No definite knowledge about
the chromopermittivity and its excess over 1 except the above conclusions is available. Moreover, the absorption may be strong enough and
hadronization (confinement) may change the final characteristics. The very first experimental indication in 1979 \cite{apan} which favored this
effect dealt with a single cosmic ray event (also some others less pronounced were observed in cosmic rays and at lower energies in emulsion
experiments at accelerators \cite{arat, alex1, alex2, masl,d90, ada1, ada2, sark1, sark2, sark3, dik, vok, gho1, gho2, deks}).

That is why it was a surprise to find in 2004 and confirm later \cite{fw1, fw2, adam, adl, ph1, ph2, ph3, ul1, ul2, pru, jia} the so-called
double-humped events in central nucleus-nucleus collisions at RHIC. In fact, the obtained distributions closely remind those published by
Cherenkov in 1937 when he used the projection of famous rings on their diameter and got the double-humped structure (e.g., see \cite{jel}). Let us
stress that this effect reflects an {\it intrinsic} property of the medium phenomenologically described by its permittivity $\epsilon $.

Experimentalists at RHIC used a quite elaborate method to learn about a parton passing through the quark-gluon medium. Imagine the large angle
scattering of two partons belonging to the colliding nuclei which happens near the surface of the medium. The triggers placed at the angle about
$\pi/2$ to the collision axis would register a normal jet created by a parton passing through the thin layer. Another scattered parton moves
through the quark-gluon medium. Both partons produce jets. The two-particle correlations were studied. Those particles which belonged to the same
jet formed a peak at low $\Delta \phi $. It was expected to get a peak in the opposite direction at $\Delta \phi \approx \pi $ due to the
correlation between particles from the trigger (near-side) and away-side jets. This was really found in pp-collisions but not in central nuclei
collisions. It implied that the away-side jet is strongly modified when passing through the quark-gluon medium. For pions produced in some energy
intervals it could not be simply attributed to jet quenching because the two-hump structure appeared. Namely this phenomenon was related to
Cherenkov gluons.

In \cite{dklv} a fit of this experimental data according to the formulae described above was performed and the values of real and imaginary parts
of the chromopermittivity were obtained. The Monte Carlo program taking into account the specifics of collisions of initial partons and
hadronization at the final stage (described by the transverse momentum spread parameter $\Delta _{\perp }$) was developed. The fit to experimental
data (with elliptic flow subtracted) is shown in Fig. 3.

\begin{figure}[ht]
 \includegraphics[width=\textwidth,height=0.5\textheight]{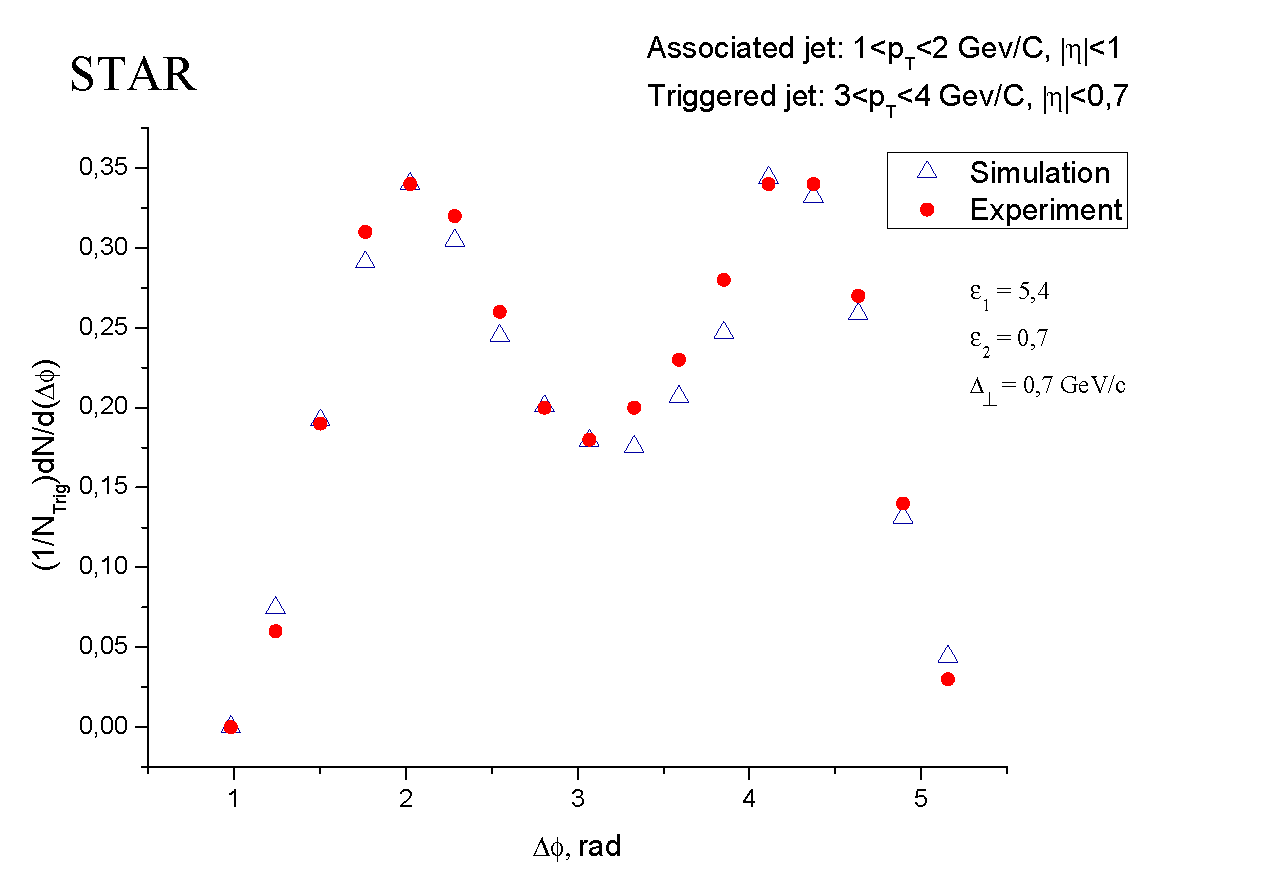}
 \caption{Associated azimuthal correlations at STAR: circles - experiment, triangles- theory.}
\end{figure}

In this way it is possible to determine the values of all the three adjustable parameters (see Table I). The constant values of $\epsilon _{1, 2}$
were used because the interval of the transferred momenta (frequencies) studied in experiment is quite narrow. Moreover, this assumption is
directly supported by experiment (for recent results see \cite{ph1, ph2, ph3, st10}) according to which the positions of peaks do not depend on
transverse momenta of both the trigger (up to $p_T^{trig}=6 $ GeV) and associated particles (up to $p_T^{assoc}=4$ GeV). That implies constant
$\epsilon_1 $ according to (\ref{9}), (\ref{xep}).

\bigskip

\begin{center}
{\bf Table 1}

\bigskip

\begin{tabular}{|c|c|c|c|c||c|}
 \hline
 Experiment & $\theta_{\rm max}$ & $\epsilon_1$ & $\epsilon_2$ &
 $\Delta_{\perp}, GeV $  & New data\\  \hline
 STAR & 1.04~rad & 5.4 & 0.7 & 0.7 & $\theta_{\rm max}\approx $1.1~rad\\
\hline
 PHENIX & 1.27~rad & 9.0 & 2.0 & 1.1& $\epsilon_1\approx 6$;  $\epsilon_2
 \approx 0.8$\\ \hline
\end{tabular}

\end{center}

\bigskip

The initial results differed for STAR and PHENIX because of disagreement in peaks positions (mostly determined by $\epsilon _1$), $\pi \pm 1.04$
and $\pi \pm 1.27$, correspondingly, and the widths (mostly determined by $\epsilon _2$). The new data \cite{st10, holz} give $\pi \pm 1.1$ so
that $\epsilon _1\approx 6$ and $\epsilon _2\approx 0.8$ would be good estimates as can be guessed from the range of values in Table 1. The main
conclusion about the large values of $\epsilon _1$ indicating on the non-gaseous matter with large $N_s$ in Eq. (\ref{f.19}) is supported in any
case. The imaginary part of the chromopermittivity is comparatively small relative to the real part. Low attenuation helps observe this effect.

The experimental data show that the often cited\footnote{E.g., "the Cherenkov angle {\it decreases} with the momentum of the radiated gluon"
\cite{E09}.} conclusion of \cite{ko} about smaller angles for larger $p_T$ is wrong. The model of a single scalar resonance used there for the
derivation of the dispersion equation defining the energy dependence of the chromopermittivity is oversimplified. This defect can be cured by the
set of overlapping resonances from PDG which leads to approximately constant $\epsilon $ in the low-energy region cut off by the Hagedorn
temperature if the formula (\ref{f.19}) is used. This set of resonances can be used to model the collective excitations of the quark-gluon medium
analogously to excited levels of atoms in electrodynamics \cite{fe}. The energy loss per 1 Fm can be estimated according to Eq. (\ref{f.17}) as
about $C_R$ GeV for the end of the resonance region near 4 GeV. It depends quadratically on this parameter that induces some flexibility in the
estimate which shows, nevertheless, that they are rather large.

For three-particle correlations \cite{holz,ajit,abel} the peak, according to the initial data, was at large angles. However, the latest data
\cite{holz} agree with the above-described estimates. These correlations show also that the hypothesis about the deflected jets \cite{salg} is not
valid because in experiment there are off-diagonal peaks which can not appear according to this hypothesis. The presence of such peaks is an
evidence for the conical emission pattern.

In-medium modifications of QCD bremsstrahlung calculations (see, e.g., \cite{vit}) predicted some broadening of the away-side peak which was,
however, not enough to get the double-humped structure after subtracting the collective flow $v_2$ effects.

In principle, the two-hump structure may arise not only due to Cherenkov gluons but also as the sonic Mach cone \cite{stoc, cas, mrup}. In
distinction to gluodynamics, it is ascribed to the {\it longitudinal} excitations in plasma \cite{mrup, rup}. The model with an hypothesized
space-like dispersion relation
\begin{equation}
\omega =\sqrt {u^2k^2+\omega _p^2}   \label{dire}
\end{equation}
was considered in \cite{rup}. Here $\omega _p$ denotes the plasma frequency,
$u$ is the speed of a plasmon. It describes a Mach shock wave for a
supersonical source. The similar structures appeared in the models using the AdS/CFT
conjecture \cite{gpy, chya}.

In the 3D hydrodynamics with big local density fluctuations which at the initial moment have the form of the tubes\footnote{Does it not remind the
field tubes in Glasma described above?}, there arise both the ridge and the double-humped structures \cite{taka, aghq, hagq}. They appear only if
the combination of these special non-smooth initial conditions with hydrodynamical evolution of the hot and dense medium is used so that the tubes
induce an appearance of the ridges in the hydrodynamical flow. At the same time, as was shown in \cite{betz}, in hydrodynamics with somewhat
different and less special initial conditions the particle yield from the {\it transverse} wake moving opposite to the trigger jet largely
overwhelms the weak Mach cone signal for relativistic particles\footnote{This conclusion differs from those in \cite{taka, aghq, hagq}, where no
away-side signal at $\Delta \phi =\pi $ is seen. Probably, this is due to the choice of a single high density spot in initial conditions of the
hydrodynamical model while two partons are necessarily created in parton treatment.}. The jet of particles in this direction should appear (the
strong away-side jet) that is not observed in experiment. As shown in section 4.3.2, this effect does not take place in chromodynamics where the
{\it longitudinal} wake (the second term in Eq. (\ref{9})) results in emission perpendicular to the parton trail. The hydrodynamical Mach cone
interpretation was not yet used for quantitative fits of experimental data aimed at getting information on the medium properties.

The charge's orientation in color space is fixed in classical equations of gluodynamics. There is no color dependence in their solutions beside
trivial Casimir factors which just determine the overall normalization. Thus the statement that "their absence ... in "conical flow" suggests {\it
sound radiation} rather than the gluonic one" \cite{shu1} is absolutely unjustified. The functional color dependence may appear in gluodynamics as
an effect at higher orders (see section 4.5).

There were several attempts to ascribe the double-humped structure around away-side jets to the large angle emission of shower gluons \cite{salg,
shu1} or to their deflection due to random multiple scattering \cite{pol} or coupling to collective flows \cite{vol, chi} but they ask for
additional ad hoc parameters with values that are not compatible with experiment \cite{ph1, ph2, ph3}. The collective medium behavior as the
origin of humps is clearly seen from properties of hadrons inside the shoulder regions ($p_T$- and centrality-independent shapes, mean $p_T$,
composition) being similar to those observed for inclusive processes \cite{ph1, ph2, ph3}. Their energy dependence suggests stronger medium
effects at higher energies, as expected in \cite{ph1, ph2, ph3}.

The method of wavelet analysis was successfully applied to distinguish between jets and conical structures in three-dimensional plots on the
event-by-event basis \cite{deks, azar}.

It was assumed everywhere above that the direction of gluons defines the direction of final pions. The new mechanism of direct emission of
Cherenkov {\it mesons} by {\it heavy} quarks was considered in the effective theory with account of the gauge/gravity duality \cite{cs}. The
Cherenkov angle is the same, but it is determined now by the relation of masses i.e. by the kinematics alone without any reference to medium
properties (the chromopermittivity).

\medskip
{\bf b. The asymmetry of in-medium resonances (SPS etc)}

The necessary condition for Cherenkov effects to be observable within some energy interval is $\epsilon (\omega )>1$ for $\omega $ belonging to
this interval. According to Eq. (\ref{f.19}) this implies that ${\rm Re} F_0(\omega )
>0$. This requirement is fulfilled within the low-energy (left) wings of
resonances described by the Breit-Wigner formula. Therefore one could expect
that the collective excitations of the quark-gluon medium leading to Cherenkov
gluons contribute in these energy intervals in addition to the traditional
resonance effects. Herefrom one gets the general prediction that the shape of
{\it any} resonance formed in nucleus collisions must become asymmetrical with
some excess within its left wing compared to usual Breit-Wigner shape. Since the
probability of Cherenkov radiation is proportional to $\Delta \epsilon =
\epsilon - 1$ this asymmetry must be proportional to $\rho (\omega )$.
Experimentally one of the best ways to observe it is to measure, say,
the dilepton mass distribution of the corresponding modes of resonance decays.
Then this distribution must have the shape \cite{dnec}
\begin{equation}
\frac{dN_{ll}}{dM}=\frac {A}{(m_{r }^2-M^2)^2+M^2\Gamma ^2}
\left(1+w_r\frac{m_{r }^2-M^2}{M^2 }\Theta (m_{r}-M)\right)    \label{ll}
\end{equation}
Here $M$ is the dilepton mass and $m_r$ is the resonance mass. The first term corresponds to the ordinary Breit-Wigner cross section with a
constant normalization factor $A$. According to the optical theorem it is proportional to the imaginary part of the forward scattering amplitude.
The second term describes the coherent Cherenkov response of the medium proportional to the real part of the amplitude. Its contribution relative
to ordinary processes is described by the only adjustable parameter $w_r$ for a given resonance $r$ which must be found from comparison with
experimental data .

This formula has been compared with the most accurate experimental results \cite{da} on the dilepton spectra of the $\rho $-resonance produced in
nuclei collisions at SPS (see Fig. 4). The asymmetry of dilepton mass spectrum is clearly seen. The corresponding parameter $w_{\rho }$ was found
to be equal to 0.19. This confirms the discussed above conclusion that the contribution of Cherenkov effects is not negligible and stresses the
fact that resonance regions may be responsible for it at rather low energies. This feature is well known in electrodynamics (see, e.g., Fig. 31-5
in \cite{fe}) where the atoms behaving as oscillators emit as Breit-Wigner resonances when get excited.

\begin{figure}[h]
\includegraphics[width=\textwidth]{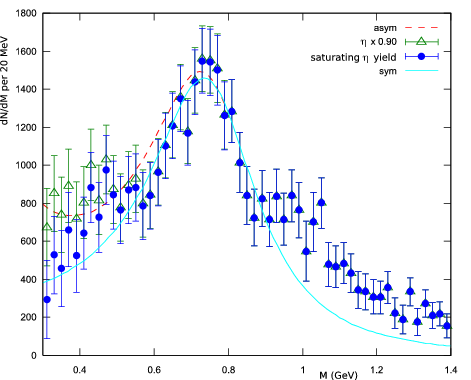}
 \caption{Modified spectrum of dileptons in semi-central collisions In(158 A GeV)-In for NA60 (points)
 compared to the $\rho $- meson peak in the medium with additional Cherenkov contribution (dashed line).}
\end{figure}

Nowadays there exist numerous data on other resonances produced in nuclei collisions with qualitative indications on asymmetry of their shapes
with excess in low-mass wings \cite{4, 5, Muto, 6, 7}. This agrees with the universal prediction of Eq. (\ref{ll}).

\medskip
{\bf c. Cosmic ray events}

The very first events with the ring-like structure were observed in non-trigger cosmic ray experiments. Namely they initiated the idea about
Cherenkov gluons \cite{d1}. Two peaks more densely populated by particles than their surroundings were noticed in the cosmic ray event \cite{apan}
initiated by a primary with energy about 10$^{16}$ eV close to LHC energies.

\begin{figure}[ht]
\includegraphics[width=\textwidth]{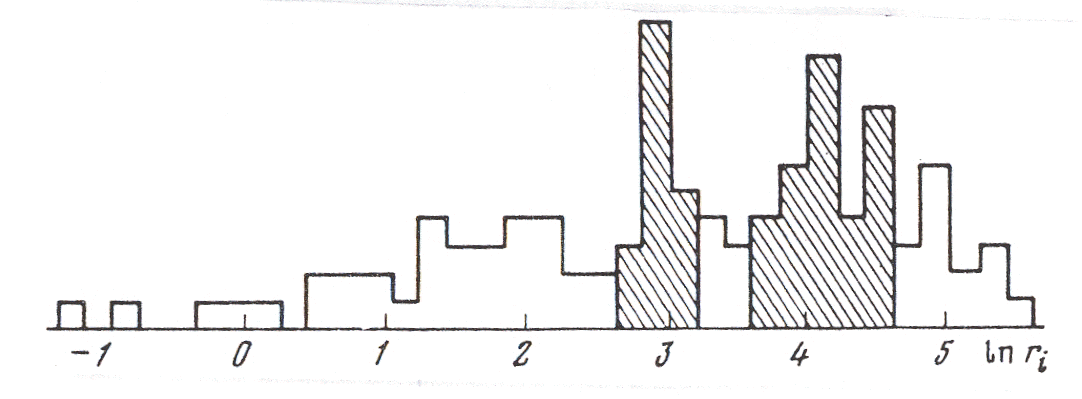}
\caption{Distribution in the number of produced particles at different distances from the collision axis $r$ in the stratospheric event at 10$^{16}$ eV
 \cite{apan} has two pronounced peaks. Correspondingly, the distribution over pseudorapidity also has two such peaks.}
 \label{coscher}
\end{figure}
\medskip

The distribution in the number of produced particles is shown in Fig. \ref{coscher},  where it is plotted as a function of the distance from the
collision axis proportional to the polar angle $\theta$. It clearly shows two maxima. This event has been registered in the detector with nuclear
and X-ray emulsions during the balloon flight at the altitude about 30 km. At the very beginning the idea about clusters (fireballs, clans, jets
etc) was proposed but it failed to explain the data. Approximately at the same time the similar event with two peaks was observed at $10^{13}$ eV
\cite{arat}. Some events with one peak (due to the limited acceptance of the installation?) were shown even earlier \cite{alex1, alex2, masl}.
When the two-dimensional distribution of particles was considered in the azimuthal plane (called as the target diagram in cosmic ray experiments),
this event revealed two (forward and backward in c.m.s.) densely populated ring-like regions within two narrow intervals of polar angles
(corresponding to peaks in Fig. \ref{coscher}) but widely distributed in azimuthal angles for each of them. Therefore such events were termed
ring-like events. The peaks were interpreted as effects due to Cherenkov gluons emitted by the forward and backward moving initial high energy
partons. In cosmic ray events the energies of partons are high and the effect can be related to the high-energy behavior of ${\rm Re}F_0(E)$.

It is important to notice that the energies of particles produced within the humps (rings) in RHIC data are rather low on the scale of initial
energies. Even the energies of jets-parents in the double-humped events are lower than 5 GeV. Therefore the whole effect at RHIC can be related to
the low-energy region.

One of the most intriguing problems is to understand properly the fact that the RHIC and cosmic ray data were fitted with very different values of
the chromopermittivity, close to 6 and 1 correspondingly. This could be interpreted as due to the difference in values of $x$ (the parton share of
energy) and $Q^2$ (the transverse momenta squared) and the Lorentz-transformation to different rest systems. It is well known that the region of
large $x$ and $Q^2$ corresponds to the dilute partonic system. At low $x$ and $Q^2$ the density of partons is much higher. Thus the effective
values of $N_s$ in (\ref{f.19}) are different.

As follows from the estimate based on Eq. (\ref{f.19}) \cite{IJMP1,IJMP2}, the density of scattering centers at RHIC conditions is very high,
about ten partons per proton volume. In this case one deals with rather low $x$ and $Q^2$. Therefore, one should expect the large density of
partons in this region and large $\Delta \epsilon $. It is interesting to note that the two-hump structure disappears in RHIC data at higher $p_T$
where the parton density gets much smaller. It corresponds to smaller $\epsilon $ and $\theta $, i.e. humps merge in the main away-side peak.

In the cosmic ray event the observed effect is related to the leading partons with large $x$. Also, the experimentalists pointed out that the
transverse momenta in this event are quite large. In this region one would expect for low parton density and very small $\Delta \epsilon $. In
distinction to a single parton traversing the medium at RHIC energies, the collision of forward moving partons at the LHC may be considered as the
collision of the two relatively dilute {\it bunches} of partons. Nevertheless, the macroscopic approach is valid there as well.

Thus the same medium can probably be considered as a liquid or a gas depending on the parton energy and transferred momenta. This statement can be
experimentally verified by using triggers positioned at different angles to the collision axis\footnote{An example is described in the next
section 4.3.2.} and considering different transverse momenta. In that way, the hadronic Cherenkov effect can be used as a tool to scan ($1/x,
Q^2$)-plane and plot on it the parton densities corresponding to its different regions.

\subsubsection{The wake}

In central collisions the interaction region is symmetrical and the change of the trigger position might lead to some effect only if the
chromopermittivity strongly depends on $\omega $. It is not likely to be the case for narrow energy intervals. However, in mid-central nuclear
collisions the interaction region itself becomes asymmetrical reminding the rugby ball. Therefore new effects can appear when the angular position
of the trigger is changed. It has been found (see Figs 2 and 3 in \cite{holz}) that the similar two-hump structure appears but with some
additional contribution noticeable at a particular orientation of the trigger particle just in between the in-plane and out-of-plane
directions\footnote{This plane is defined by the collision axis and the impact parameter.}. It was stated in \cite{holz} that "at present, it is
unclear whether this merely reflects a geometry dependent shift in the away-side peaks or perhaps an additional contribution at $\Delta \phi =\pi
/2$".

We have provided arguments \cite{wake} that this is an additional contribution in the vicinity of $\Delta \phi = \pi /2$ known in electrodynamics
as the wake effect (see, e.g., \cite{ryaz}). At the same time, it is strongly influenced by the special geometry of mid-central collisions in such
a way that the wake radiation at about $\pi /4$-orientation of the trigger particle can escape the overlap region much easier than wake gluons for
in- and out-of-plane orientations of the trigger particle.

According to Eq. (\ref{9}) the second term corresponds to the radiation induced by the longitudinal oscillations in the wake left behind the
away-side parton. It is emitted mostly in the direction perpendicular to the wake. That reminds the dipole radiation. Namely the color
charge-changing regions of the dipole-like structure behind the parton are seen in the Monte Carlo simulations of the wake \cite{cmt, cmrt}
trailing the color parton. The enhanced and depleted charge density regions alternate. The $1/\sqrt x$-singularity is however stronger than the
$\sin ^2\theta $ dipole distributions. It must be somehow saturated in the vicinity of $x=0$ by the spatial dispersion of $\epsilon $. Therefore
one may use this term for qualitative conclusions in the region not very close to $x=0$.

The estimates \cite{wake} of  the relative importance of the two terms in Eq. (\ref{9}) show that the wake radiation is negligible at the peak of
the Cherenkov humps at $x_0$. However, they become comparable at
\begin{equation}
x_e\approx \frac {x_0^2}{2x_0+1}
\end{equation}
that corresponds to the angles $\pi-\Delta \phi _L\approx 1.43 $ rad. The wake radiation overwhelms the Cherenkov contribution at $\Delta \phi
_L<\pi -1.43$ rad where the latter decreases. Therefore the shift of their combined maximum to $\pi-\Delta \phi _L\approx 1.3$ rad noticed in
\cite{holz} and shown in Fig. 2 in this paper seems quite plausible if explained as the effect of the wake.

Let us note the difference between Cherenkov and wake radiations. Sometimes the name of wake is attributed to both of them \cite{rup,mrup}.
However, it can be shown \cite{wake,ryaz} that the trail behind the parton (the wake) induced by the chromopermittivity corresponds just to the
second term in (\ref{9}). Moreover, the wake radiation exists only for non-zero imaginary part of $\epsilon _l$ while the Cherenkov radiation also
exists and gets the traditional $\delta $-functional angular shape for real $\epsilon $. The effect of transverse wake in hydrodynamics is
completely different. It produces a strong peak in the away-side direction at $\Delta \phi_L = \pi$ \cite{betz} which is not observed in
experiment.

\subsubsection{Transition radiation}

In 1945 V.L. Ginzburg and I.M. Frank proposed the idea about the transition radiation \cite{gfra}. Any electric charge passing from some medium to
another one or moving with (almost) constant velocity in inhomogeneous medium with variable dielectric permittivity induces its polarization and
radiation of photons. This property has been used, e.g., to build the transition radiation tracker in ATLAS (A Toroidal LHC ApparatuS) detector at
the Large Hadron Collider (LHC). The transition radiation is determined by restructuring of the electromagnetic field surrounding the electric
charge.

During hadron (nucleus) collisions the partons intersect the surface of the partner particle (nucleus). This transition should induce abrupt
changes of their color fields and, consequently, radiation of gluons. The classical in-medium QCD equations coincide in their general structure
with electrodynamical equations. Since this approach happens to be successful with Cherenkov gluons and wakes it is tempting to apply the
Ginzburg-Frank formula to estimate the number of gluons emitted per unit length within some frequency and angular intervals
\begin{eqnarray}
\frac {dN}{d\omega d\Omega } =
\frac {\alpha _S \sqrt {\epsilon _a} \sin^22\theta}{\pi \omega }\cdot
\hspace{8cm}  \nonumber \\
\cdot \left \vert \frac {(\epsilon _b - \epsilon _a)(1-\epsilon _a+\sqrt {\epsilon _b - \epsilon _a \sin ^2\theta})}{(1-\epsilon _a\cos
^2\theta)(1+\sqrt {\epsilon _b - \epsilon _a \sin ^2\theta})(\epsilon _b\cos \theta + \sqrt {\epsilon _a\epsilon _b-\epsilon _a^2 \sin
^2\theta})}\right \vert ^2, \label{gfra}
\end{eqnarray}
where $\epsilon _{a, b}$ are the chromopermittivities of the media $a, b$ as felt by relativistic ($v\approx c$) partons traversing the surface
when passing from $b$ to $a$.  In equation (\ref{gfra}) we have omitted the color Casimir factors.

There is important difference between Cherenkov and transition radiations. Cherenkov radiation can be observed only if the particle moves in the
medium with $\epsilon >1$. Its intensity is proportional to $\epsilon -1$ according to (\ref{f.17}). Transition radiation is proportional to
$(\epsilon _b - \epsilon _a)^2$ (see (\ref{gfra})) and appears at any change of $\epsilon $. In electrodynamics it is studied mostly in the
frequency interval where the formula (\ref{eed}) is applicable i.e. where $\Delta \epsilon <0$. Namely this behavior of the dielectric
permittivity is responsible for specific features of the transition radiation. In the case of the nuclear medium it is unknown whether the
analogous formula may be used in any energy region. For narrow energy intervals it is reasonable to assume $\Delta \epsilon $ to be constant. At
very high energies the dependence (\ref{eom}) may be used. Both these assumptions find some support in RHIC and cosmic ray data as discussed
above. That is why we consider \cite{trt} first these two possibilities.

The away-side parton at RHIC is created inside the quark-gluon medium, moves through it and escapes in the vacuum where it hadronizes. According
to the above estimates $\epsilon _b \approx 6\gg \epsilon _a=1$. Using (\ref{gfra}) one gets the bremsstrahlung-like spectrum with both infrared
and collinear divergencies in asymptotics
\begin{equation}
dN\propto \frac {d\omega }{\omega }\frac {d\theta }
{\theta }.     \label{sle}
\end{equation}
At finite energies the angular divergence is replaced by the maximum at $\theta \approx \gamma ^{-1}$. Anyway, this radiation is collimated near
the initial direction of the away-side parton and is hard to detect because of the strong background. The similar conclusions are obtained
\cite{kleo} for the radiation by a gluon moving in a stochastic medium. It is interesting to note that for $\epsilon _b \gg \epsilon _a$ the
intensity of the transition radiation (\ref{sle}) does not depend on $\epsilon _b$.

According to (\ref{eom}) for the forward moving partons with extremely high energies the deviation of chromopermittivity from unity is very small.
Then
\begin{equation}
dN\propto \frac {d\omega }{\omega ^5}\frac {d\theta }
{\theta }\Theta (\omega -\omega _{th}).     \label{she}
\end{equation}
The collinear divergence remains the same but the energy spectrum gets a strong peak near $\omega _{th}$. It implies that the resonance-like
almost monochromatic subjets with masses close to $\omega _{th}$ will be produced at LHC energies. These "resonances" are mostly created by gluons
and, therefore, are neutral. The most favored channel would be to observe them as peaks in mass spectra of forward moving $\mu ^+\mu ^-$-pairs
near $\omega _{th}$. Unfortunately, there exists no clear prescription how to translate the threshold energy in hadronic reactions to gluons. If
observed, peaks locations and their disappearence at $\omega = \omega _{th}$ due to $\Theta (\omega -\omega _{th})$ would provide us information
about that.

One may hope that the confinement and hadronization do not spoil these conclusions. This is supported by observation of effects due to Cherenkov
gluons. However, the transition radiation might be stronger influenced by confinement near the QGP surface than collective excitations inside the
volume. From one side, it shortens the radiation length but, from another side, the abrupt variation of color fields surrounding partons when they
come out of the quark-gluon plasma and hadronize may enlarge the transition radiation because of strong confining forces. At the same time the
transition radiation may appear in much wider energy intervals than Cherenkov radiation (in particular, at $\omega < \omega _{th}$). However,
neither experimental nor theoretical arguments about the chromopermittivity in these regions exist nowadays. It would not be surprising if we find
its effects outside the regions where the double-humped events were observed. That may somewhat smoothen down the structures near $\omega _{th}$.

Another way to account for difference of vacuum and medium chromopermittivities is to consider different masses of gluons and values of the
running coupling in these two enviroments. It was done in \cite{zakh, dgyu1, dgyu2} where the transition radiation of heavy quarks was treated in
microscopic approach. It was shown that it can be as important as the induced gluon radiation. However, the conclusions are still indefinite
because of strong cancellation effects and assumptions about the behavior of the coupling constant.

One should also point out important problem of the finite size $L$ of hadronic objects. It always induces (see, e.g., \cite{ zakh, djor, d0,
tamm}) the factors like $1-\cos L/L_0$ in emission probabilities. Estimates of the scale length $L_0$ are still not very definite. This problem is
closely connected with decisions about the radiation (formation) lengths and/or about the points where the partons were created.

\subsection{Instabilities at high energies}

The a'la tachyonic model (\ref{eom}) proposed for the chromopermittivity at high energies is inevitably related with instabilities of the
quark-gluon medium. The dielectric permittivity of the macroscopic matter (e.g., as given by (\ref{eed})) is usually considered in its rest system
which is well defined. For collisions of two nuclei (or hadrons) such a system requires special definition (see \cite{inmed, okun}). In
particular, for fast forward moving partons the spectators (the medium) are formed by the partons of another (target) nucleus at rest. Thus we
consider a problem of the system of the quark-gluon medium impinged by a bunch of fast partons similar to the problem of plasma physics, namely,
that of the interaction of the electron bunch with plasma \cite{rukh,term,aruk}. This differs from RHIC conditions when partons scattered at $\pi
/2$ were considered as moving in the medium with its rest system coinciding with the nuclei center of mass system.

Two complications are to be considered.
First, there are selected directions of the current ($z$-axis) and its radiation.
This is cured by introducing the permittivity tensor as
\begin{equation}
D_i(\omega , {\bf k})=\epsilon _{ij}(\omega , {\bf k})E_j(\omega , {\bf k}).
\end{equation}
Second, the impinging bunch content is similar to that of the target
(plasma-plasma collisions!) and its permittivity must also be accounted for.
In the projectile system it is the same as the permittivity of the target in
its rest system, i.e. given by (\ref{eom}). Then one takes into account that
the total induced current is the sum of currents induced in the target and in
the projectile and Lorentz-transforms the projectile internal fields,
polarization vector and currents to the target rest system where we consider
the whole process. In this way the Lorentz transformation of the
conductivity tensor $\sigma _{i, j}(\omega , {\bf k})$
(and, correspondingly, of $\epsilon _{ij}(\omega , {\bf k}$)) is found as it
is done in \cite{rukh, term, aruk, sruk, ginz}. Using
the current conservation and classical field equations (\ref{6}), (\ref{7})
with $g=0$ one gets for non-zero components of the chromopermittivity tensor
\begin{eqnarray}
\epsilon_{xx}=\epsilon_{yy}=1+\frac {\omega_0^2}{\omega ^2}(1+\frac {1}{\gamma}),
\nonumber \\
\epsilon _{xz}=-\epsilon_{zx}=\frac {\omega_0^2k_T}{\omega ^2(\omega -k_z)\gamma },
\nonumber \\
\epsilon _{zz}=1+\frac {\omega_0^2}{\omega ^2}\left(1+\frac {k_T^2}{(\omega -k_z)^2
\gamma }\right ).                    \label{tens}
\end{eqnarray}
Here $k_T$ and $k_z$ are the transverse and longitudinal components of
${\bf k}$. We use the approximation of high energies (large $\gamma \gg 1 $
factor, i.e. $v\approx 1$). The terms depending on $\gamma $ are due to the
impinging partons. They can be omitted everywhere except the terms which
determine the Cherenkov gluon radiation at $\omega -k_z\approx 0$.

The classical equations derived from (\ref{f.6}), (\ref{f.7}) and written
in the momentum space have solution if the following dispersion equation
is valid
\begin{equation}
{\rm det}(\omega, {\bf k})=\vert k^2\delta _{ij}-k_ik_j-\omega ^2\epsilon _{ij}
\vert =0.  \label{disp}
\end{equation}
It is of the sixth order in momenta dimension. However, the sixth order terms
cancel and (\ref{disp}) leads to two equations (of the second order):
\begin{equation}
k^2-\omega ^2-\omega _0^2=0,     \label{plas}
\end{equation}
\begin{equation}
(k^2-\omega ^2-\omega _0^2)(1+\frac {\omega_0^2}{\omega^2})-
\frac {\omega _0^4k_T^2}{\omega ^2(\omega-k_z)^2\gamma }=0.  \label{bunch}
\end{equation}
They determine the internal modes of the medium and the bunch propagation
through the medium, correspondingly.

The equation (\ref{plas}) shows that the quark-gluon medium is unstable because there exists the branch with ${\rm Im} \; \omega >0$ for modes
$k^2<\omega _0^2$. Thus the energy increase of the total cross sections responsible for positiveness of ${\rm Re} \; F_0(\omega )$ at high
energies is related to the instability of the quark-gluon medium.

The equation (\ref{bunch}) has solutions corresponding to Cherenkov gluons emitted by the impinging bunch and determined by the last term in
(\ref{bunch}). They can be found at $\omega=k_z+\delta \; (\delta \ll \omega)$. For $k_T=\omega _0$ one gets the solutions with
\begin{equation}
{\rm Im} \; \delta_1=\frac {\omega _0^2}{2k_z[2\gamma (1+\omega_0^2/k_z^2)]^{1/3}}. \label{d1}
\end{equation}
For $k_T\neq \omega_0$ there is the solution with
\begin{equation}
{\rm Im} \; \delta_2=\frac {\omega _0^2k_T}{k_z[\gamma \vert k_T^2-\omega _0^2\vert (1+\omega_0^2/k_z^2)]^{1/2}}.                 \label{d2}
\end{equation}
It is well known (see \cite{kruk}) that the solutions of the disperion
equation (\ref{disp}) determine the Green function of the system equations
\begin{equation}
G(t,z)=\frac {1}{2\pi^2}\int _{-\infty}^{\infty}dk\int _{C(\omega )}\frac {1}
{{\rm det}(\omega, {\bf k})}\exp (-i\omega t+ikz)d\omega,      \label{green}
\end{equation}
where the contour $C(\omega )$ passes above all singularities in the integral. Therefore, the positive ${\rm Im} \; \delta _i $ in (\ref{d1}) and
(\ref{d2}) correspond to the absolute instability of the system. Let us note that the instability at $k_T=\omega _0$ is stronger than at $k_T\neq
\omega _0$ approximately by the factor $\gamma ^{1/6}$ (this factor is about 4 times larger at LHC compared to RHIC). The instability exponent
(\ref{d1}) decreases as $\gamma ^{-1/3}$ and is about 16 times smaller at LHC compared to RHIC. It tends to zero asymptotically.

Thus Cherenkov gluons are emitted with constant transverse momentum $k_T=\omega _0$ and their number,at high energies, is proportional to
$$
\frac{d\omega}{(\omega )^2} \; \Theta (\omega -\omega _{th}),
$$
where $\epsilon (\omega )$ is given by Eq. (\ref{eom}) with account of the threshold above which this equation is applicable. It differs from the
traditional folklore of constant emission angle of Cherenkov radiation and the number of gluons $\propto d\omega $ (or the total energy loss
proportional to $\omega d\omega $). This difference is easily explained by Eqs (\ref{f.10}), (\ref{f.17}) which give $\cos \theta $=const and the
energy loss $\omega d\omega $ for $\epsilon $=const and $k_T\approx \omega _0$ and $d\omega /\omega $ energy loss for $\epsilon = 1+\omega
_0^2/\omega ^2$.

If the data about ${\rm Re} \; F_0(\omega )$ for hadronic processes are used in (\ref{f.19}) and fitted by (\ref{eom}) then the value of $\omega
_0$ depends on the density of scatterers $N_s$ and with above estimates can be of GeV order.

\subsection{Nonlinear effects and the color rainbow}

In the equations of in-medium QCD (\ref{f.6}),(\ref{f.7}) considered in the subsection 4.1 one assumed the chromopermittivity to be abelian, $\epsilon^{ab} = \delta^{ab} \epsilon$. The simplest generalization of the equations (\ref{f.6}),(\ref{f.7}) taking into account the nonabelian properties of the chromopermittivity $\epsilon^{ab}$ take into account the tensor structure of $\epsilon^{ab}$. Let us write the corresponding generalization of (\ref{f.6}),(\ref{f.7}) in the leading order in the coupling constant:
\begin{eqnarray}
 \bigtriangleup {\bf A}^a - \epsilon^{ab} \frac{\partial ^2 {\bf A}^b}
 {\partial t^2} &=&  - {\bf j}^a, \nonumber \\
 \epsilon^{ab}(\bigtriangleup \Phi^b - \epsilon^{bc} \frac{\partial ^2}
 {\partial t^2} \Phi^c) &=& - \rho^a,
 \label{teneps}
\end{eqnarray}
where $(\rho^a,{\bf j}^a )$ are the components of the external color current $j^a_\mu$ defined in (\ref{f.11}).

From the physical meaning of the chromopermittivity $\epsilon^{ab}$ there follows \footnote{In the present subsection we consider the simplest case of the real chromopermittivity.} that the most general expression for the chromopermittivity is a symmetric tensor $\epsilon^{ab}=\epsilon^{ba}$ with the equal diagonal $\epsilon^{({\rm d})}$ and off-diagonal $\epsilon^{({\rm o})}$ components. To elucidate the physical content of the equations (\ref{teneps}) let us consider the similarity transformation $U$ diagonalizing the matrix $\epsilon^{ab}$:
\begin{equation}\label{diag}
\epsilon \;\; \rightarrow \;\; {\tilde \epsilon}=U \epsilon U^{-1}.
\end{equation}
In (\ref{diag}) the matrix ${\tilde \epsilon}$ is the diagonal matrix with the eigenvalues of the chromopermittivity tensor $\{ \epsilon^{(a)} \}$, where $a=1, \cdots, N^2_c-1 $, on the diagonal. It is easy to show that within the above-described assumptions on the structure of $\epsilon^{ab}$ one has:
\begin{eqnarray}\label{eigen}
\epsilon^{(1)} & \equiv & \epsilon^* =\epsilon^{({\rm d})}+
2\epsilon^{({\rm o})}, \nonumber \\
\epsilon^{(2), \cdots, (N^2_c-1)} & \equiv & \epsilon^{**}=
\epsilon^{({\rm d})}-\epsilon^{({\rm o})}.
\end{eqnarray}
The rotation $U$ allows to diagonalize (\ref{teneps}) in terms of the modified fields ${\tilde A}^a_\mu=U^{ab} A^b_\mu$ and the modified current
${\tilde j}^a_\mu=U^{ab} j^b_\mu$:
\begin{eqnarray}
 \bigtriangleup {\tilde {\bf A}}^a - \epsilon^{(a)} \frac{\partial ^2 }
 {\partial t^2} {\tilde {\bf A}}^a &=&  - {\tilde {\bf j}}^a, \nonumber \\
 \bigtriangleup {\tilde \Phi}^a - \epsilon^{(a)} \frac{\partial ^2}
 {\partial t^2} {\tilde \Phi}^a &=& - \frac{{\tilde \rho}^a}{\epsilon^{(a)}}.
 \label{tenepsd}
\end{eqnarray}
From the above-described structure of the eigenvalues (\ref{eigen}) there follows an existence of the two different solutions of the equations (\ref{tenepsd}) corresponding to the eigenvalues $(\epsilon^*,\epsilon^{**})$:
\begin{eqnarray}\label{modsol}
{\tilde \Phi}^{(1)\;1} ({\bf r},t)&=&\frac {2g}{\epsilon^*} \frac {\Theta
(vt-z-r_{\perp }\sqrt {\epsilon^* v^2-1})}{\sqrt {(vt-z)^2-r_{\perp} ^2
(\epsilon^* v^2-1)}}, \nonumber \\
{\tilde \Phi} ^{(1) \; 2, \cdots, N^2_c-1}({\bf r},t)&=&\frac {2g}{\epsilon^{**} }\frac {\Theta
(vt-z-r_{\perp }\sqrt {\epsilon^{**} v^2-1})}{\sqrt {(vt-z)^2-r_{\perp} ^2
(\epsilon^{**} v^2-1)}},
\end{eqnarray}
and the analogous solutions for ${\tilde {\bf A}}^a$. Returning to the original fields $A^a_\mu=\left(U^{-1} {\tilde A} \right)^a_\mu$ one evidently gets the solution corresponding to the presence of the color Cherenkov rainbow formed by the Cherenkov radiation at the angles
\begin{equation}
\label{mcos}
\cos \theta^{*} = \frac {1}{v \sqrt {\epsilon^{({\rm d})}+2\epsilon^{({\rm o})}}}, \;\;\;\;
\cos \theta^{**} = \frac {1}{v \sqrt {\epsilon^{({\rm d})}-
\epsilon^{({\rm o})}}}.
\end{equation}
In the limit $\epsilon^{({\rm o})} \ll \epsilon^{({\rm d})}$ the solution (\ref{modsol}) turns into the standard solution (\ref{f.11}),(\ref{f.12}).

To account for the nonlinear contributions to the in-medium Yang-Mills equations it is necessary to use the corresponding generalizations of
(\ref{teneps}). To demonstrate the possible practical consequences of such an approach in \cite{okun} one described the procedure of the model
calculations of the nonlinear contributions to the gluon Cherenkov radiation, albeit with some simplified (and nonrealistic) assumptions
concerning the color structure of the chromopermittivity. It was shown that in the case under consideration in the solutions there appear
contributions of the form
$$
\propto \frac{\Gamma _t {\sqrt x}}{[(x-x_0)^2+(\Gamma _t)^2/4]^{5/4}}
$$
and
$$
\propto \frac{\Gamma _t } {{\sqrt x}[(x-x_0)^2+(\Gamma _t)^2/4]^{3/4}},
$$
which correspond to the color rainbow and differ (despite some similarity) from the lowest order solution (\ref{9}).

\subsection{Hydrodynamics (thermodynamical and mechanical properties of QGP)}

The chromodynamical properties of QGP described above are its average characteristics measured by a test parton. Therefore they are more robust
than the thermodynamical and mechanical characteristics which change during the evolution of the quark-gluon medium. Spatial and temporal
information is necessary to get the latter. The lattice calculations and ideas about CGC, Glasma, QGP provide some hints to transition from
confined state to deconfined quarks and gluons. The inelastic collisions may lead to thermalization of the medium. Subsequently, it expands and
hydrodynamics may be applied to treat this stage of evolution. It is described in many reviews (e.g., see \cite{sh1, hein}) and we give just a
brief survey of it here.

The medium is characterized by 6 independent variables. Those are the energy density $e$, pressure $p$, baryon number $n_B$ and 3 components of
the velocity vector $u_{\mu }$. The energy-momentum tensor and baryon number current are
\begin{equation}
T^{\mu \nu }(x)=(e(x)+p(x))u^{\mu }(x)u^{\nu }(x)-p(x)g^{\mu \nu },
\;\;\;  j_B^{\mu }(x)=n_B(x)u^{\mu }(x).     \label{tj}
\end{equation}
The evolution of these variables is described by six equations of hydrodynamics. Those are five non-linear partial differential equations obtained
from local conservation laws for energy, momentum and baryon number
\begin{equation}
\partial _{\mu } T^{\mu \nu }(x)=0 \;\; (\nu = 0, ..., 3);
\;\;\;  \partial _{\mu }j_B^{\mu }(x)=0     \label{ctj}
\end{equation}
and an equation of state relating $p,\;e$ and $n_B$. The latter is usually chosen in the closest possible form to the results of lattice QCD by
normalization on the states below and above the critical temperature (i.e., hadrons to a quark-gluon medium). This is a rather arbitrary element
of the whole approach. Besides, the solutions of the non-linear equations in (3+1)-dimensions ask for initial conditions to be defined and may be
only obtained numerically with several external parameters. That is the origin of several conflicting results as was discussed above with
references to \cite{rup, stoc, cas, mrup, taka, aghq, betz}. The CGC and Glasma approaches provide some guesses to the initial conditions for the
evolution of a thermalized QGP. However, the evolution of the Glasma into a thermalized QGP is not yet understood.

At least four parameters are necessary to fix the initial conditions and freeze-out algorithm which determines the transition from the
hydrodynamical characteristics to hadronic stage. Those are the initial time $\tau_{eq}$, energy (or entropy) density $s_{eq}$, baryon number
$n_{B,eq}$ and the freeze-out temperature (or decoupling energy density) $e_{dec}$. They are fixed by comparison of the theoretical results with
the experimental data. Schematically, the correspondence of these parameters and experimental characteristics can be represented as $dN/dp_T - T -
e_{dec}; \; dN/dy - (\tau s)_{eq}; \; p/\pi - n_{B,eq}/s_{eq}; \; (dN/dp_T)_p/(dN/dp_T)_{\pi } - \tau _{eq}$. Moreover, from the overlap geometry
of the two colliding nuclei one can define the shape of the initial density profiles as well as make assumptions concerning the profiles of the
initial longitudinal and transverse flows and the prescription for final hadronization.

From solutions of the hydrodynamical equations one gets the transverse momentum spectra for various species of particles, the radial and elliptic
flows, the shape of the interaction region (as revealed by Bose-Einstein correlations and Hanburry Brown-Twiss interferometry). Their comparison
with experimental data allows to determine the main thermodynamical, statistical and mechanical properties of the quark-gluon medium.
Symbolically, their values found from 200 GeV data can be grouped as $T_{eq}\approx 360$ MeV; $T_{cr}\approx 170$ MeV; $T_{dec}\approx 120$ MeV;
$\tau _{therm}=\tau _{eq}\approx 0.6<1$ fm; $\tau _{dec}\approx 7$ fm; $e_{th}\approx 25$ GeV/fm$^3$; $e_{cr}\approx 1$ GeV/fm$^3$;
$e_{dec}\approx 0.075$ GeV/fm$^3$; $s_{eq}\approx 110$ fm$^{-3}$; $\eta /s\approx 0.1$ ($1/4\pi $ in AdS/CFT); $n_B<0.5$ fm$^{-3}$. These values
show rapid thermalization, high average initial energy density and quite long "lifetime" of the quark-gluon plasma before hadronization at rather
low energy density and temperature predicted by lattice QCD. Complete thermalization in the time less than 1 fm is required to obtain the measured
value of elliptic flow and its centrality dependence which is very sensitive to any deviation from local thermal equilibrium of low $p_T$
particles. Collective excitations, resonances and inelastic collisions keep the system in thermal equilibrium. The good agreement of the data with
ideal fluid dynamics points to a very small viscosity of QGP. Other transport coefficients (shear, diffusion, heat conduction) are not important
if the microscale defined by rescattering is much less than the macroscale related to medium expansion. Strong non-perturbative interaction should
be responsible for its behavior as an ideal liquid. Herefrom comes the name of strong quark-gluon plasma (sQGP). In sQGP, there may exist, e.g.,
clusters \cite{yuk} and colored bound states of massive quasiparticles \cite{szah} with heavier flavors. Resonance scattering on constituents of
the quark-gluon medium can become important \cite{rapp}. All that would give rise to its collective response with new scales discussed above, high
pressure, large chromopermittivity and long-range correlations necessary to explain the enhancement of strange partons production by non-local
processes. The long "lifetime" supports approximate description of energy losses within an "infinite" medium.

Hydrodynamics is an actively developing field nowadays. Main characteristics of low $p_T$ particles have been described in this approach. At the
same time many factors have to be taken into account such as thermodynamics, models of collective flows, hadronization process, resonance decays,
chemical composition, geometry of collisions etc. However a fully consistent hydrodynamical description has not yet been found. Some controversial
conclusions appear from time to time about, e.g., the energy (from SPS to RHIC), rapidity and centrality dependence of elliptic flow, its absolute
value, the transverse momentum dependence of various radii derived from HBT analysis, the chemical composition of some species (e.g., $\bar
{p}/\pi $ ratio) etc. One may hope that they will be resolved in the near future within the same set of adjusted parameters. That would provide
deeper understanding of collective thermodynamical and mechanical properties of the bulk matter and its evolution.

\section{Some new possibilities at the LHC}

First LHC experiments on pp interactions at 2.36 and 7 TeV already showed that even the most general characteristics like inclusive pseudorapidity
distributions in the central region $\vert \Delta \eta \vert < 2.4$ for CMS (Compact Muon Solenoid) and $\vert \Delta \eta \vert < 1$ for ALICE (A
Large Ion Collider Experiment) are not well enough described by theoretical predictions. Even though the qualitative features are reproduced, no
Monte Carlo scheme (among the four tried) fits the results quantitatively. Surely, more surprises are waiting for us in correlation
studies\footnote{The CMS Collaboration reported \cite{ppridge} an observation (!) of a ridge with the narrow distribution in relative azimuthal
angle and broad distribution in pseudorapidity difference in pp collisions at the energy of 7 TeV.}, especially, in heavy-ion experiments. The
energies of LHC open new possibilities for studies of the reviewed topics. The gluon-gluon processes will dominate. The formation length for
radiation of gluons (and photons) becomes larger at higher energies. Thus high energy partons will feel the finite size of the interaction region.
The spectra of radiated gluons and photons are predicted to become harder due to oscillations of the light-cone wave function \cite{Z04, AZ09}.
The emission from heavy quarks should be enhanced by the finite-size effects relative to light quarks. In general, the relative roles of surface
effects, initial high virtualities of created high-$p_T$ partons, dead-cone effect in radiation by heavy quarks will be elucidated. Coherent final
state interactions must become stronger.

The smaller values of chromopermittivity for very high energy partons discussed above indicates on weaker coupling and lower parton density of the
medium at LHC compared with RHIC. Therefore, the role of fluctuations induced by instabilities will be enhanced at LHC \cite{mr08}.

Both trigger and non-trigger experiments are important in AA-collisions. Following RHIC in trigger experiments, the partons (jets) at large angles
with much higher energies will be available. Correspondingly, this allows to study jet quenching, ridge and double-humped events in wider energy
intervals. In particular, the energy dependence of the chromopermittivity may be found by studying humps behavior in wider energy intervals. The
humps shifts due to the wake effect in semi-central collisions and due to Lorentz transformation of the chromopermittivity tensor in a wide energy
interval are of a special interest.

In non-trigger experiments studies of the response of the quark-gluon medium to the extremely fast moving partons will be of primary importance.
It would be quite instructive if the events similar to the cosmic ray one are observed. The finite-size effects are especially important for
forward-moving partons with large $x$.

The search for forward subjets with a fixed mass (due to the transition radiation - Eq. (\ref{she})) or fixed transverse momentum (due to high
energy Cherenkov gluons - Eq. (\ref{bunch})) might be fruitful.

Effects due to Cherenkov gluons can be used to study the wide region of $(x, Q^2)$-plane available at the LHC. The different values of parton
densities would correspond to different subregions on this plane.

The nonlinear effects described in section 4.5 may become more pronounced albeit still difficult to separate from main effect due to the possible
dispersive behavior of the chromopermittivity.

One can imagine more exotic experiments. For example, let triggers register a single electron (positron) which is produced in a process of the
convolution of the quark-antiquark pair inverse to those studied at $e^+e^-$-colliders. The dielectric permittivity can be measured and the
formulae (\ref{eed}), (\ref{eom}) compared with the data. Even though the cross section is very small, it might be measurable in the high
luminosity regime of LHC.

\section{Conclusions}

The experimental studies at SPS and, especially, the numerous and detailed data about heavy ion collisions at RHIC provided us with invaluable
information about the completely new field -- the properties of the quark-gluon medium. They showed that AA events can not be described as an
additive collection of pp collisions. The heavy ion dynamics must take into account the collective behavior of the medium. That is clearly
demonstrated by anisotropic flows, jet quenching, such peculiar correlations as ridge and double-humped events etc.

Theoretical understanding of medium evolution asked for application of QCD at full strength to speculate about such stages as CGC, Glasma,
thermalization, QGP, hadronization. The specific configuration of fields in Glasma and corollaries of deconfinement at QGP stage need further
studies. In its turn, experimental results required to revive the theoretical methods widely applied in the condensed matter physics. The
modification of energy losses of partons in the medium because of the formation length effects and its collective response (the
chromopermittivity) due to polarization are described in this manner. Hydrodynamics was extensively exploited for description of collective
behavior of this medium.

Both experimental and theoretical progress in this field are extremely fruitful and promising. Surely, even more interesting results will be
obtained in further searches at RHIC. Exciting perspectives are opened with the advent of LHC in the new energy region.
 \medskip

{\bf \large Acknowledgements}

\medskip

This work was supported by RFBR grants 09-02-00741; 08-02-91000-CERN and by the RAN-CERN program.\\

\medskip

{\bf Figure captions.}                                  \\

 Fig. 1.  QCD cascade in the medium.\\

 Fig. 2.  Distribution in rapidity $P(y)$ of final prehadrons: 1) $L=0\,{\rm fm}$, full angular ordering, red, full line; 2) $L=0.5\,{\rm fm}$,
 partial angular ordering, green, dashed line; 3) $L=5\,{\rm fm}$, partial angular ordering, blue, dotted line \cite{LN10}.\\

 Fig. 3.  Associated azimuthal correlations at STAR: experiment- circles, theory- triangles. \\

 Fig. 4. Modified spectrum of dileptons in semi-central collisions In(158 A GeV)-In for NA60 (points)
 compared to the $\rho $-meson peak in the medium with additional Cherenkov contribution (dashed line).\\

 ���. 5. Distribution in the number of produced particles at different distances from the collision axis $r$ in the stratospheric event at 10$^{16}$ eV
 \cite{apan} has two pronounced peaks. Correspondingly, the distribution in pseudorapidity also has two such peaks. \\


\begin{thebibliography}{99}
\bibitem{CGC1}
Iancu E, Venugopalan R arXiv:hep-ph/0303204
\bibitem{CGC2}
Venugopalan R arXiv:hep-ph/0412396
\bibitem{CGC3}
Leonidov A {\it Phys. Usp.} {\bf 48} 323 (2005)
\bibitem{CGC4}
Gelis F, Iancu E, Jalilian-Marian J, Venugopalan R, arXiv:1002.0333
\bibitem{M03}
McLerran L D arXiv:hep-ph/0311028
\bibitem{MV1}
McLerran L D, Venugopalan R {\it Phys. Rev. D} {\bf 49} 2233 (1994)
\bibitem{MV2}
McLerran L D, Venugopalan R {\it Phys. Rev. D} {\bf 49} 3352 (1994)
\bibitem{MV3}
McLerran L D, Venugopalan R {\it Phys. Rev. D} {\bf 50} 2225 (1994)
\bibitem{JV04}
Jeon S, Venugopalan R {\it Phys.Rev. D} {\bf 70} 105012 (2004)
\bibitem{L08}
Lappi T {\it Eur. Phys. J. �}\; {\bf 55} 285 (2008)
\bibitem{KMW95}
Kovner A, McLerran L D, Weigert H {\it Phys. Rev. D}\; {\bf 52} 62231 (1995)
\bibitem{FKL06}
Fries R J, Kapusta J I, Li Y arXiv:nucl-th/0604054
\bibitem{LM06}
Lappi T, McLerran L {\it Nucl. Phys. A}\; {\bf 772} 200 (2006)
\bibitem{L06}
Lappi T {\it Phys. Lett. B}\; {\bf 643} 11 (2006)
\bibitem{V05}
Venugopalan R arXiv:hep-ph/0412396
\bibitem{MMR98}
Matinyan S G, Muller B, Rischke D H {\it Phys. Rev. C}\; {\bf 57} 1927 (1998)
\bibitem{FGM07}
Fukushima K, Gelis F, McLerran L {\it Nucl. Phys. A}\; {\bf 786} 107 (2007)
\bibitem{MS04}
Mueller A H, Son D T {\it Phys. Lett. B} {\bf 582} 279 (2004)
\bibitem{RV061}
Romatschke P, Venugopalan R {\it Phys. Rev. Lett.}\; {\bf 96} 06232 (2006)
\bibitem{RV062}
Romatschke P, Venugopalan R {\it Phys. Rev. D}\; {\bf 74} 045011 (2006)
\bibitem{FI08}
Fujii H, Itakura K {\it Nucl. Phys. A}\;  {\bf 809} 88 (2008)
\bibitem{mr881}
Mrowczynski S {\it Phys. Lett. B}\; {\bf 214} 587 (1988); {\bf 314} 2197 (1993)
\bibitem{mr882}
Mrowczynski S {\it Phys. Lett. B}\; {\bf 314} 2197 (1993)
\bibitem{ALM03}
Arnold P, Lenaghan J, Moore G D {\it JHEP}\;{\bf  08} 002 (2003)
\bibitem{BMSS01}
Baier R, Mueller A H, Schiff D, Son D T {\it Phys. Lett. B}\;  {\bf 502} 51 (2001)
\bibitem{AM06}
Arnold P, Moore G D {\it Phys. Rev. D}\; {\bf 73}  025006 (2006)
\bibitem{MSW07}
Mueller A H, Soshi A I, Wong S M H {\it Nucl. Phys. B}\; {\bf 760} 145 (2007)
\bibitem{K08}
Khachatryan V, {\it Nucl. Phys. A}\; {\bf 810} 109 (2008)
\bibitem{JIMWLK1}
Jalilian-Marian J, Kovner A, Leonidov A, Weigert H {\it Phys. Rev. D} {\bf 59} 034007 (1999)
\bibitem{JIMWLK2}
Iancu E, Leonidov A, McLerran L D {\it Nucl. Phys. A} {\bf 692} 583 (2001)
\bibitem{JIMWLK3}
Ferreiro E, Iancu E, Leonidov A, McLerran L D {\it Nucl. Phys. A} {\bf 703} 489 (2002)
\bibitem{W02}
Weigert H {\it Nucl. Phys. A}\; {\bf 703} 823 (2002)
\bibitem{BK1}
Balitsky I {\it Nucl. Phys. B}\; {\bf 463} 99 (1996)
\bibitem{BK2}
Kovchegov Yu. V. {\it Phys. Rev. D}\; {\bf 61} 074018 (2000)
\bibitem{HEF1}
Gelis F, Lappi T, Venugopalan R {\it Phys. Rev. D}\; {\bf 78} 054019 (2008)
\bibitem{HEF2}
Gelis F, Lappi T, Venugopalan R {\it Phys. Rev. D}\; {\bf 78} 054020 (2008)
\bibitem{HEF3}
Gelis F, Lappi T, Venugopalan R {\it Phys. Rev. D}\; {\bf 79} 094017 (2009)
\bibitem{E09}
d'Enterria D arXiv:0902.2011 [nucl-ex]
\bibitem{BSZ00}
Baier R, Schiff D, Zakharov B G {\it Ann. Rev. Nucl. Part. Sci.}\; {\bf 50} 37 (2000)
\bibitem{KW03}
Kovner A, Wiedemann U A "Gluon radiation and parton energy loss", in R.C. Hwa and X.-N. Wang (eds) "Quark Gluon Plasma 3" 192 (2003)
\bibitem{GVWZ03}
Gyulassy M, Vitev I, Wang X-N, Zhang B-W "Jet quenching and radiative energy loss in dense nuclear matter", in R.C. Hwa and X.-N. Wang (eds)
"Quark Gluon Plasma 3", 123 (2003)
\bibitem{Z04a}
Zakharov B G  arXiv:hep-ph/0412.117
\bibitem{SS07}
Casalderrey-Solana J, Salgado C A {\it Acta Phys. Polonica B}\; {\bf 38} 3731 (2007)
\bibitem{W09}
Wiedemann U A arXiv:0908.2306 [hep-ph]
\bibitem{M10}
Majumder A arXiv:1002.2206 [hep-ph]
\bibitem{DGLAP1}
Gribov V N, Lipatov L N {\it Sov. Journ. Nucl. Phys.}\; {\bf 15} 438 (1972)
\bibitem{DGLAP2}
Altarelli G, Parisi G {\it Nucl. Phys. B}\; {\bf 126} 298 (1977)
\bibitem{DGLAP3}
Dokshitzer Yu L {\it Sov.Phys.  JETP}\; {\bf 46} 641 (1977)
\bibitem{coscad}
Belenky S Z  {\it Shower Processes in Cosmic Rays}, Moscow, Gostechizdat, 1948 (in Russian)
\bibitem{pionizationNAO1}
Dremin I M {\it JETP Lett.}\, {\bf 31} 185 (1980)
\bibitem{pionizationNAO2}
Dremin I M, Leonidov A V {\it Sov. Phys. Usp.}\, {\bf 23} 515 (1980)
\bibitem{pionizationNAO3}
Dremin I M, Leonidov A V {\it Sov. Journ. Nucl. Phys.}\, {\bf 35} 247 (1982)
\bibitem{pionizationNAO4}
Dremin I M, Leonidov A V {\it Sov. Phys. Usp.}\, {\bf 38} 723 (1995)
\bibitem{pionizationAO1}
Leonidov A V, Ostrovsky D M  {\it Phys. Atom. Nucl.}\, {\bf 60} 110 (1997)
\bibitem{pionizationAO2}
Leonidov A V, Ostrovsky D M, {\it Phys. Atom. Nucl.}\, {\bf 62} 701 (1999)
\bibitem{Z07}
Zakharov B G {\it JETP Lett.}\; {\bf 86} 444 (2007)
\bibitem{BDMPS1}
Baier R et al. {\it Phys. Lett. B}\; {\bf 345} 277 (1995)
\bibitem{BDMPS2}
Baier R et al. {\it Nucl. Phys. B}\; {\bf 483} 291 (1997)
\bibitem{BDMPS3}
Baier R et al. {\it Nucl. Phys. B}\; {\bf 484} 265 (1997)
\bibitem{BDMPS4}
Baier R et al. {\it Phys. Rev. C}\; {\bf 58} 1706 (1998)
\bibitem{GLV1}
Gyulassy M Levai P, Vitev I {\it Nucl. Phys. B}\; {\bf 571} 197 (2000)
\bibitem{GLV2}
Gyulassy M Levai P, Vitev I {\it Phys. Rev. Lett.}\; {\bf 85} 5535 (2000)
\bibitem{GLV3}
Gyulassy M Levai P, Vitev I {\it Nucl. Phys. B}\; {\bf 94} 371 (2001) (2005)
\bibitem{GLV4}
Gyulassy M Levai P, Vitev I {\it Phys. Lett. B}\; {\bf 538} 282 (2002)
\bibitem{GLV5}
Djordjevic M, Gyulassy M, {\it Nucl. Phys. A}\; {\bf 733} 265 (2004)
\bibitem{GLV6}
Djordjevic M, Gyulassy M, Wocks S, {\it Phys. Rev. Lett.}\; {\bf 94} 112301 (2005)
\bibitem{W00}
Wiedemann U A {\it Nucl. Phys. B}\; {\bf 588} 303 (2000)
\bibitem{ASW1}
Wiedemann U A {\it Nucl. Phys. B}\; {\bf 582} 409 (2000);
\bibitem{ASW2}
Wiedemann U A {\it Nucl. Phys. A}\; {\bf 690} 731 (2001)
\bibitem{ASW3}
Salgado C A, Wiedemann U A {\it Phys. Rev. D}\; {\bf 68} 014008 (2003)
\bibitem{ASW4}
Armesto N, Salgado C A, Wiedemann U A {\it Phys. Rev. D}\; {\bf 69} 114003 (2004)
\bibitem{Z96}
Zakharov B G {\it JETP Lett}\; {\bf 63} 952 (1996)
\bibitem{Z1}
Zakharov B G {\it JETP Lett}\; {\bf 65} 615 (1997)
\bibitem{Z2}
Zakharov B G {\it Phys. Atom. Nucl.}\; {\bf 61} 838 (1998)
\bibitem{Z99}
Zakharov B G {\it JETP Lett.}\; {\bf 70} 176 (1999)
\bibitem{HT1}
Guo X F, Wang X N {\it Phys. Rev. Lett.}\; {\bf 85} 3591 (2000)
\bibitem{HT2}
Majumder A, Muller B {\it Phys. Rev. C}\; {\bf 77} 054903 (2008)
\bibitem{AMY1}
Arnold P, Moore G D, Yaffe L G {\it JHEP}\; {\bf 11} 057 (2001)
\bibitem{AMY2}
Arnold P, Moore G D, Yaffe L G {\it JHEP}\; {\bf 12} 009 (2001)
\bibitem{AMY3}
Arnold P, Moore G D, Yaffe L G {\it JHEP}\; {\bf 06} 030 (2002)
\bibitem{AMY4}
Turbide S et al. {\it Phys. Rev. C}\; {\bf 72} 014906 (2005)
\bibitem{AMY5}
Qin G Y et al. {\it Phys. Rev. Lett.}; {\bf 100} 072301 (2008)
\bibitem{KP00}
Kampfer B, Pavlenko O {\it Phys. Lett.}\; {\bf B477} 171 (2000)
\bibitem{Z87}
Zakharov B G {Sov. Journ. Nucl Phys.}\; {\bf 46} 92 (1987)
\bibitem{SP09}
Peigne S, Smilga A {\it Phys. Usp.}\; {\bf 52} 659 (2009)
\bibitem{Z04}
Zakharov B G {\it JETP Lett.} {\bf 80} 1 (2004)
\bibitem{ACSX08}
Armesto N, Cunqueiro L, Salgado C, Xiang W-C {\it JHEP}\; {\bf 0802} 048 (8008)
\bibitem{WHS96}
Wang X N, Huang Z, Sarcevic I {\it Phys. Rev. Lett.}\; {\bf 77} 457 (2006)
\bibitem{PYTHIA}
Sjostrand T, Mrenna S, Skands P {\it JHEP}\ {\bf 0605} 026 (2006)
\bibitem{HERWIG1}
Marchesini G, Webber B R, Abbiendi G, Knowles I G, Seymour M H, Stanco L {\it Comput. Phys. Commun.}\; {\bf 67} 465 (1992)
\bibitem{HERWIG2}
Corcella G, Knowles I G, Marchesini G, Moretti S, Odagiri K, Richardson P, Seymour M H, Webber B {\it JHEP}\; {\bf 0101} 010 (2001)
\bibitem{HERWIG3}
Corcella G, Knowles I G, Marchesini G, Moretti S, Odagiri K, Richardson P, Seymour M H, Webber B hep-ph/0210213
\bibitem{PYQUEN}
Lokhtin I P, Snigirev A M {\it Eur. Phys. J. C}\; {\bf 45} 211 (2006)
\bibitem{JEWEL}
Zapp K, Ingelman G, Rathsman J, Stachel J, Wiedemann U A {\it Eur. Phys. J C} {\bf 60} 617 (2009)
\bibitem{QPYTHIA}
Armesto N, Cunqueiro L, Salgado C A {\it Eur. Phys. J C} {\bf 63} 679 (2009)
\bibitem{QHERWIG}
Armesto N, Corcella G, Cunqueiro L, Salgado C A {\it JHEP} {\bf 0911} 122 (2009)
\bibitem{LN10}
Leonidov A V, Nechitailo V A, arXiv:1006.0366 [nucl-th]
\bibitem{B84}
G. Bergmann, {\it Phys. Rep.}\; {\bf 107} 1 (1984)
\bibitem{CS86}
Chakravarty S, Schmid A, {\it Phys. Rep.}\; {\bf 140} 195 (1986)
\bibitem{SG1}
Selikhov A V, Gyulassy M {\it Phys. Lett. B}\ {\bf 316} 373 (1993)
\bibitem{SG2}
Selikhov A V, Gyulassy M {\it Phys. Rev. C}\ {\bf 49} 1726 (1994)
\bibitem{MCLPM1}
Zapp K, Stachel I, Wiedemann U A {\it Phys. Rev. Lett.}\; {\bf 103} 152302 (2009)
\bibitem{MCLPM2}
Zapp K, Stachel I, Wiedemann U A  {\it Nucl. Phys. A}\; {\bf 830} 171 (2009)
\bibitem{WG01}
Wang X N, Guo X F {\it Nucl. Phys. A}\; {\bf 696} 788 (2001)
\bibitem{BW05}
Borghini N, Wiedemann U A, arXiv:hep-ph/0506218
\bibitem{LMPSAT09}
Lokhtin I P, Malinina L V, Petrushanko S V, Snigirev A M, Arsene I, Tywoniuk K {\it Comput. Phys. Commun.}\; {\bf 180} 779 (2009)
\bibitem{DS06}
Dremin I M, Shadrin O S {\it J. Phys. G}\; {\bf 32} 963 (2006)
\bibitem{APQ09}
Armesto N, Pajares C, Quiroga-Arias P, {\it Eur. Phys. J. C}\; {\bf 61} 779 (2009)









\bibitem{sh1}
Shuryak E V {\it Prog. Part. Nucl. Phys.} {\bf 62} 48 (2009)
\bibitem{hein}
Heinz U arXiv:hep-ph/0407360; CERN-2004-001
\bibitem{inmed}
Dremin I M {\it Eur. Phys. J. C} {\bf 56} 81 (2008)
\bibitem{dpri}
Djongolov M K, Pisov S, Rizov V {\it J. Phys. G} {\bf 30} 425 (2004)
\bibitem{ja}
Jackson J D {\it Classical electrodynamics} John Wiley and Sons Inc.,
 Fig. 7.9 (1998)
\bibitem{fe}
Feynman R P, Leighton R B, Sands M  {\it The Feynman Lectures in Physics} Addison-Wesley PC Inc, vol.~1, ch.~31 (1963)
\bibitem{kk}
Kalashnikov O K, Klimov V V {\it Sov. J. Nucl. Phys.} {\bf 31} 699 (1980)
\bibitem{kl1}
Klimov V V {\it Sov. J. Nucl. Phys.} {\bf 33} 934 (1981)
\bibitem{kl2}
Klimov V V {\it Sov. Phys. JETP} {\bf 55} 199 (1982)
\bibitem{we}
Weldon H A {\it Phys. Rev. D} {\bf 26} 1394 (1982)
\bibitem{bi}
Blaizot J P, Iancu E {\it Phys. Rep.} {\bf 359} 355 (2002)
\bibitem{rrs1}
Rebhan A, Romantschke P, Strickland M {\it Phys. Rev. Lett.} {\bf 94} 102303 (2005)
\bibitem{rrs2}
Rebhan A, Romantschke P, Strickland M  {\it JHEP} {\bf 0509} 041 (2005)
\bibitem{amy}
Arnold P, Moore G D, Yaffe L G  {\it Phys. Rev. D} {\bf 72} 054003 (2005)
\bibitem{qmsh}
Qin G Y, Majumder A, Song H, Heinz U {\it Phys. Rev. Lett.} {\bf 103} 152303 (2009)
\bibitem{ko}
Koch V, Majumder A, Wang X N {\it Phys. Rev. Lett.} {\bf 96} 172302 (2006)
\bibitem{rup}
Ruppert J {\it J. Phys. Conf. Ser.} {\bf 27} 217 (2005)
\bibitem{zakh}
Zakharov B G {\it JETP Lett.} {\bf 76} 201 (2002)
\bibitem{cs}
Casalderrey-Solana J, Fernandez D, Mateos D arXiv:0912.3717
\bibitem{dgyu1}
Djordjevic M, Gyulassy M {\it Phys. Rev. C} {\bf 68} 034914 (2003)
\bibitem{dgyu2}
Djordjevic M, Gyulassy M {\it Phys. Lett. B} {\bf 560} 37 (2003)
\bibitem{djor}
Djordjevic M {\it Phys. Rev. C} {\bf 73} 044912 (2006)
\bibitem{go}
Goldberger M, Watson K {\it Collision Theory} John Wiley and Sons Inc., Ch. 11, sect.~3, sect.~4 (1964)
\bibitem{scad}
Scadron M D {\it Advanced quantum theory and its applications through Feynman diagrams} Springer-Verlag, p. 326 (1979)
\bibitem{d1}
Dremin I M {\it JETP Lett.} {\bf 30} 140 (1979)
\bibitem{d0}
Dremin I M {\it Sov. J. Nucl. Phys.} {\bf 33} 726 (1981)
\bibitem{bol}
Bolotovsky B M, Stolyarov S N {\it Sov. Phys. Uspekhi} {\bf 17} 875 (1975)
\bibitem{alf}
Alfimov M arXiv:1004.0286
\bibitem{apan}
Apanasenko A V et al. {\it JETP Lett.} {\bf 30} 145 (1979)
\bibitem{kruk}
Kuzelev M V, Rukhadze A A {\it Metody teorii voln v sredah s dispersiey} M Fizmatlit (2007); {\it Methods of wave theory in dispersive media}
Singapore WSPC (2009)
\bibitem{tfra}
Tamm I E, Frank I M {\it Dokl. Akad. Nauk SSSR} {\bf 14} 107 (1937)
\bibitem{gr}
Grichine V M {\it Nucl. Instr. Meth. A} {\bf 482} 629 (2002)
\bibitem{dklv}
Dremin I M, Kirakosyan M R, Leonidov A V, Vinogradov A V {\it Nucl. Phys. A} {\bf 826} 190 (2009)
\bibitem{wake}
Dremin I M {\it Mod. Phys. Lett. A} {\bf 25} 591 (2010)
\bibitem{tgyu}
Thoma M H, Gyulassy M {\it Nucl. Phys. B} {\bf 351} 491 (1991)
\bibitem{arat}
Arata N {\it Nuovo Cim. A} {\bf 43} 455 (1978)
\bibitem{alex1}
Alexeeva K I et al. {\it Izv. AN SSSR} {\bf 26} 572 (1962)
\bibitem{alex2}
Alexeeva K I et al. {\it J. Phys. Soc. Japan} {\bf 17}, A-II (1962)
\bibitem{masl}
Maslennikova N V et al. {\it Izv. AN SSSR} {\bf 36} 1696 (1972)
\bibitem{d90}
Dremin I M et al {\it Sov. J. Nucl. Phys.} {\bf 52} 536 (1990)
\bibitem{ada1}
Adamovich M I et al {\it J. Phys. G} {\bf 19} 2035 (1993)
\bibitem{ada2}
Adamovich M I et al  {\it Eur. Phys. J. A} {\bf 5} 429 (1999)
\bibitem{sark1}
Gogiberidze G L et al {\it Phys. Lett. B} {\bf 430} 368 (1998)
\bibitem{sark2}
Gogiberidze G L et al {\it Phys. Lett. B} {\bf 471} 257 (1999)
\bibitem{sark3}
Gogiberidze G L et al {\it Phys. Atom Nucl.} {\bf 64} 143 (2001)
\bibitem{dik}
Dremin I M, Ivanov O V, Kalinin S A et al {\it Phys. Lett. B} {\bf 499} 97 (2001)
\bibitem{vok}
Vokal S et al {\it Phys. Atom. Nucl.} {\bf 71} 1395 (2008)
\bibitem{gho1}
Ghosh D et al {\it Can. J. Phys.} {\bf 86} 919 (2008)
\bibitem{gho2}
Ghosh D et al {\it Ind. J. Phys.} {\bf 82} 1339 (2008)
\bibitem{fw1}
Wang F (STAR Collab.) {\it J. Phys. G} {\bf 30} S1299 (2004)
\bibitem{fw2}
Wang F (STAR Collab.) {\it J. Phys. G} {\bf 34} S337 (2007)
\bibitem{adam}
Adams J et al (STAR Collab.) {\it Phys. Rev. Lett.} {\bf 95} 152301 (2005)
\bibitem{adl}
Adler S S et al. (PHENIX Collab.) {\it Phys. Rev. Lett.} {\bf 97} 052301 (2006)
\bibitem{ph1}
Adare A et al. (PHENIX Collab.) {\it Phys. Rev. Lett.} {\bf 101} 232301 (2008)
\bibitem{ph2}
Adare A et al. (PHENIX Collab.)  {\it Phys. Rev. C} {\bf 77} 011901 (2008)
\bibitem{ph3}
Adare A et al. (PHENIX Collab.)  {\it Phys. Rev. C} {\bf 78} 014901 (2008)
\bibitem{ul1}
Ulery J G (STAR Collab.) {\it J. Phys. G} {\bf 35} 104032 (2008)
\bibitem{ul2}
Ulery J G (STAR Collab.){\it Int. J. Mod. Phys. E} {\bf 16} 3123 (2008); arXiv:0709.1633; 0801.4904
\bibitem{pru}
Pruneau C A {\it Phys. Rev. C} {\bf 79} 044907 (2009)
\bibitem{jia}
Jiangyong Jia {\it J. Phys. G} {\bf 35} 104033 (2008)
\bibitem{jel}
Jelley J V {\it Cherenkov radiation and its applications} Pergamon Press p. 13 (1958)
\bibitem{st10}
Aggarwal M M et al (STAR Collab) arXiv:1004.2377
\bibitem{holz}
Holzmann W G (PHENIX Collab.) {\it Nucl. Phys. A} {\bf 830} 781c (2009)
\bibitem{ajit}
Ajitanand N N (PHENIX Collab.) {\it Nucl. Phys. A} {\bf 783} 519 (2007)
\bibitem{abel}
Abelev B I et al. (PHENIX Collab.) {\it Phys. Rev. Lett.} {\bf 102} 052302 (2009); arXiv:0912.3977
\bibitem{salg}
Salgado C {\it J. Phys. G} {\bf 35} 054001 (2008)
\bibitem{vit}
Vitev I {\it Phys. Lett. B} {\bf 630} 78 (2005)
\bibitem{stoc}
St\"ocker H {\it Nucl. Phys. A} {\bf 750} 121 (2005)
\bibitem{cas}
Casalderrey-Solana J, Shuryak E V, Teaney D {\it J. Phys. Conf. Ser.} {\bf 27} 22 (2005)
\bibitem{mrup}
Ruppert J, M\"uller B {\it Phys. Lett. B} {\bf 618} 123 (2005)
\bibitem{gpy}
Gubser S S, Pufu S S, Yarom A {\it JHEP} {\bf 09} 108 (2007)
\bibitem{chya}
Chesler P M, Yaffe L G {\it Phys. Rev. Lett.} {\bf 99} 152001 (2007)
\bibitem{taka}
Takahashi J et al {\it Phys. Rev. Lett.} {\bf 103} 242301 (2009)
\bibitem{aghq}
Andrade B P G, Grassi F, Hama Y, Quian W-L arXiv:0912.0703
\bibitem{hagq}
Hama Y, Andrade B P G, Grassi F, Quian W-L arXiv:0911.0811
\bibitem{betz}
Betz B et al. {\it Phys. Rev. C} {\bf 79} 034902 (2009)
\bibitem{shu1}
Shuryak E V {\it J. Phys. G} {\bf 35} 104044 (2008)
\bibitem{pol}
Polosa A D, Salgado C A {\it Phys. Rev. C} {\bf 75} 041901 (2007)
\bibitem{vol}
Voloshin S A {\it Nucl. Phys. A} {\bf 749} 287 (2005)
\bibitem{chi}
Chiu C, Hwa R {\it Phys. Rev. C} {\bf 74} 064909 (2006)
\bibitem{deks}
Dremin I M, Eyyubova E G, Korotkikh V L, Sarycheva L I {\it J. Phys. G} {\bf 35} 095106 (2008)
\bibitem{azar}
Azarkin M {\it Diploma work} MEPhI (2010)
\bibitem{dnec}
Dremin I M, Nechitailo V A {\it Int. J. Mod. Phys. A} {\bf 24} 1221 (2009)
\bibitem{da}
Damjanovic S et al. (NA60 Collab.) {\it Phys. Rev. Lett.} {\bf 96} 162302 (2006)
\bibitem{4}
Trnka D et al. {\it Phys. Rev. Lett.} {\bf 94} 192303 (2005)
\bibitem{5}
Naruki M et al. {\it Phys. Rev. Lett.} {\bf 96} 092301 (2006)
\bibitem{Muto}
Muto R et al. {\it Phys. Rev. Lett.} {\bf 98} 042501 (2007)
\bibitem{6}
Kozlov A (PHENIX Collab.) {\it Eur. Phys. J. A} {\bf 31} 836 (2007)
\bibitem{7}
Kotulla M (CBELSA/TAPS Collab.)  {\it AIPConf. Proc.} {\bf 870} 506 (2006)
\bibitem{IJMP1}
Dremin I M {\it J. Phys. G} {\bf 35} 054001 (2008)
\bibitem{IJMP2}
Dremin I M {\it Int. J. Mod. Phys. A} {\bf 22} 3087 (2007)
\bibitem{ryaz}
Ryazanov M I {\it Elektrodynamika condensirovannyh sred} M. Nauka (1984) pp. 210-215
\bibitem{cmt}
Chakraborty P, Mustafa M G, Thoma M H {\it Phys. Rev. D} {\bf 74}
 094002 (2004)
\bibitem{cmrt}
Chakraborty P, Mustafa M G, Ray R, Thoma M H {\it J. Phys. G} {\bf 34} 214 (2007)
\bibitem{gfra}
Ginzburg V L, Frank I M {\it J. Phys. USSR} {\bf 9} 353 (1945); {\it ZhETF} {\bf 16} 15 (1946)
\bibitem{trt}
Dremin I M arXiv:1003.2145
\bibitem{kleo}
Kirakosyan M R, Leonidov A V arXiv:0810.5442
\bibitem{tamm}
Tamm I E {\it J. Phys. USSR} {\bf 1} 439 (1939)
\bibitem{okun}
Dremin I M {\it Yad. Fiz.} {\bf 73} 684 (2010)
\bibitem{rukh}
Rukhadze A A {\it ZhTF} {\bf 62} 669 (1962)
\bibitem{term}
Ter-Mikaelyan M L {\it High-energy electromagnetic processes in condensed matter} NY Wiley-Interscience (1972)
\bibitem{aruk}
Alexandrov A F, Rukhadze A A {\it Lekzii po electrodynamike plasmopodobnyh sred; neravnovesnye sredy} M Fizicheskii facultet MGU, p. 112 (2002)
\bibitem{sruk}
Silin V P, Rukhadze A A {\it Electromagnitnye svoystva plasmy i plasmopodobnyh sred} M. Gosatomizdat, p. 173 (1961)
\bibitem{ginz}
Ginzburg V L {\it Rasprostranenie electromagnitnyh voln v plasme} M. Nauka (1967); {\it The propagation of electromagnetic waves in plasmas}
Oxford New York, Pergamon Press (1970)
\bibitem{yuk}
Yukalov V I, Yukalova E P {\it Physica A} {\bf 243} 382 (1997)
\bibitem{szah}
Shuryak E V, Zahed I {\it Phys. Rev. D} {\bf 70} 054507 (2004)
\bibitem{rapp}
Rapp R et al {\it Nucl. Phys. A} {\bf 830} 861c (2009)
\bibitem{AZ09}
Aurenche P, Zakharov B G {\it JETP Lett.} {\bf 90} 237 (2009)
\bibitem{mr08}
Mrowczynski S {\it Acta Phys. Pol. B} {\bf 39} 941 (2008); arXiv:0804.0275
\bibitem{ppridge}
Khachatryan V et al. (CMS Collab) JHEP (09) 091 (2010)

\end{thebibliography}
\end{document}